\documentclass[%
 aip,
 amsmath,amssymb,
 reprint,%
]{revtex4-2}

\usepackage{graphicx}
\usepackage{dcolumn}
\usepackage{tikz}
\usepackage{bm}

\usepackage[utf8]{inputenc}
\usepackage[T1]{fontenc}
\usepackage{mathptmx}
\usepackage{etoolbox}

\makeatletter
\def\@email#1#2{%
 \endgroup
 \patchcmd{\titleblock@produce}
  {\frontmatter@RRAPformat}
  {\frontmatter@RRAPformat{\produce@RRAP{*#1\href{mailto:#2}{#2}}}\frontmatter@RRAPformat}
  {}{}
}%
\makeatother
\begin{document}

\preprint{AIP/xxx-QED}

\title[Phenomenological NLO modeling in thin films]{Modeling nonlinear optical interactions of focused beams in bulk crystals and thin films: A phenomenological approach}

\author{Kai J. Spychala}
\affiliation{Department of Physics, University of Paderborn, Warburger Str. 100, 33098 Paderborn, Germany}
\author{Zeeshan H. Amber}
\affiliation{TU Dresden, Institute of Applied Physics, Nöthnitzer Strasse 61, 01187 Dresden, Germany}

\author{Lukas M. Eng}
\affiliation{TU Dresden, Institute of Applied Physics, Nöthnitzer Strasse 61, 01187 Dresden, Germany}
\affiliation{ct.qmat: Dresden-Würzburg Cluster of Excellence—EXC 2147, TU Dresden, 01062 Dresden, Germany}

\author{Michael Rüsing}
\affiliation{TU Dresden, Institute of Applied Physics, Nöthnitzer Strasse 61, 01187 Dresden, Germany}

\email{Correspondence:michael.ruesing@tu-dresden.de}

\date{\today}

\begin{abstract}
Coherent nonlinear optical $\mu$-spectroscopy is a frequently used tool in modern material science, as it is sensitive to many different local observables, which comprise, among others, crystal symmetry and vibrational properties.
The richness in information, however, may come with challenges in data interpretation, as one has to disentangle the many different effects like multiple reflections, phase jumps at interfaces, or the influence of the Guoy-phase.
In order to facilitate interpretation, the work presented here proposes an easy-to-use semi-analytical modeling ansatz, that bases upon known analytical solutions using Gaussian beams. Specifically, we apply this ansatz to compute nonlinear optical responses of (thin film) optical materials. We try to conserve the meaning of intuitive parameters like the Gouy-phase and the nonlinear coherent interaction length. In particular, the concept of coherence length is extended, which is a must when using focal beams. The model is subsequently applied to exemplary cases of second-harmonic and third-harmonic generation.
We observe a very good agreement with experimental data and furthermore, despite the constraints and limits of the analytical ansatz, our model performs similarly well as when using 
more rigorous simulations. However, it outperforms the latter in terms of computational power, requiring more than three orders less computational time and less performant computer systems.  
\end{abstract}

\maketitle

\section{\label{sec:Introduction}Introduction}
The development of potent and reliable nonlinear optical materials enables a broad field of nonlinear optical (NLO) applications \cite{Yuan2021}, such as frequency converters\cite{BS17}, sources of squeezed light, as well as heralded single photons\cite{OBJ12}, to name a few examples. The materials need to fulfill the typical demands concerning optical materials, i.e. showing low absorption/scattering, and a reliable quality. These criteria are supplemented by the requirement of large optical non-linearities. Furthermore, it is very important to be able to create functional structures, for example waveguide-networks, or to combine the material for instance with silicon, to form hybrid structures\cite{MORU19,MX19}. In typical material systems like lithium niobate (LN), the established technologies of waveguide formation and other functional structures get increasingly replaced by the utilization of thin films (tf) and other nanosized structures \cite{Yuan2021,Liu22, tien1971light}.\\ 
One of the key tasks in the material development and device technology is quality control\cite{pulker1979characterization} and improvement. Here parametric nonlinear optical processes allow for a non-invasive, fast, and very sensitive probing of different properties. 
Therefore, they are widely used as a material characterization technique in form of second-harmonic- and third-harmonic-microscopy\cite{Spy2020FM,HERBU17, ZRus21,squier1998third, shen1989surface, LZWX20,HRu22-2} as well as via Coherent Anti Stokes Raman Scattering\cite{Reitzig:22,hempel2021broadband} (CARS), for example.\\%
When characterizing thin films and using (strong) focusing, the typical textbook description of the nonlinear optical interaction, however, breaks down rapidly, or is at least inaccurate. One main issue is for example, that results are often interpreted based on plane wave models, which do not include focusing. This, for example, leads to a massive overestimation of the coherent interaction length. To counter this, many numerical simulation codes are available\cite{Sandkuijl2013} and deliver good results, but are often not able to provide results that intuitively allow for disentangling different effects, while also needing significant computing resources\cite{ZRus21}. This limits the ability to scan large parameter spaces.
Therefore, this paper tries to clarify the possible applications and limits of the analytical textbook results. Furthermore, we aim to extent these approaches to focused beams and interpret those solutions. Here, we present a modeling toolkit which enables accurate description for typical characterization setups of (thin film) nonlinear optical materials. At the same time we aim at conserving and/or extending the established nomenclature and interpretation of typical parameters like the coherent interaction length, the confocal parameter and the Gouy-phase in order to preserve a simple and intuitive understanding. This approach saves computation power and can help in the development of new methods or rapid anlysis of results by allowing quick identification of key parameters in experiments.\par 
This paper is structured in six sections. The sections do not necessarily need to be read in consecutive order as they address different interests such as looking for modeling results for specific processes, discussions of the details of the modeling or its limits and prospects.\\
If the reader is interested in a brief overview of established textbook descriptions, on which our semi-analytical model is built on, and a discussion on their limitations, this can be found in Sec.~II. This section presents a brief reminder on textbook description of nonlinear optical interactions in crystals. Here, the plane wave and Gaussian beam approaches are discussed and the relevant parameters are introduced. \\ 
Section~III. contains a short discussion on the general relations and limits between full numerical and semi-analytical solutions in the context of nonlinear interactions in focused beams.\\
The improved semi-analytical approach is introduced in Sec.~IV. This chapter deals with the extension and reformulation of the standard analytical textbook models in order to reliably describe nonlinear optical microscopy in thin films. Crucially, Sec.~IV~B discusses the role and meaning of the coherent interaction length in focused beams and its differences compared to the well established theory that bases on plane waves.\\

Should the reader be mainly interested in the performance of the model versus experimental data, as well as its power for predicting nonlinear optical phenomena in focused beams, then the reader may directly jump to Sec.~V. This section contains different applications of the model for bulk crystal second-harmonic generation (SHG), 
as well as SHG and third-harmonic generation (THG) in thin films.\\

Sec.~VI closes the paper with a summarizing discussion and outlook.

\section{Textbook approaches and their limits} \label{sec:tbres}
 In many lectures on nonlinear optics, the model calculations are introduced on the basis of planar waves, cf. Eq.~\eqref{eq:plwave}, with a position-dependent amplitude $a_q$ as solutions for the nonlinear wave equation:
 \begin{equation}\label{eq:plwave}
E(z,t)=a_q(z)\cdot(e^{i(k_qz-\omega_q t)}+c.c.)\;.
 \end{equation}
 Here, $q$ represents the harmonic order, i.e. $q=2$ belongs to SHG. One then makes use of the slowly varying envelope approximation and solves the approximated version of the nonlinear wave equation, so that the amplitude $a$ of the newly generated wave at a specific position, e.g. where it leaves the crystal, can be computed via Eq.~\eqref{eq:plane-waves-ansatz}. It yields the archetype $sinc(x)$ solution as:
 \begin{equation}
     a\propto \int e^{i\Delta k z} dz \;.
     \label{eq:plane-waves-ansatz}
 \end{equation}
 Here, the phase mismatch $\Delta \vec{k}= \sum_{i,o}\vec{k}_i-\vec{k}_o $, i.e. the difference between wave-vectors of the incoming ($\vec{k}_i$) and produced waves ($\vec{k}_o$), is the key parameter one needs to know in order to predict any signal.\\
This picture may already deliver a good description if its approximations are fulfilled, i.e. there is almost no focusing. In the textbook of Boyd \cite{Boyd92}, the paraxial coupled-wave-equations, cf. Eq.~\eqref{eq:coupledwaveeq}, are solved using a Gaussian ansatz as in Eq.~\eqref{eq:gauss-ansatzqz}, and therefore include also weak to medium focusing conditions:
\begin{align}
    & 2ik_q \frac{\partial A_q}{\partial z}+\nabla_T^2A_q=-\frac{\omega_q^2}{\epsilon_0 c^2} P_q e^{(i\Delta k z)} \label{eq:coupledwaveeq}\; ;\\
       & A(z,r)=\frac{a_q(z)}{1+\frac{i2z}{b}} e^{\left(-\frac{q r^2}{w_0^2(1+i2z/b)}\right)} e^{i(k_qz-\omega_q t)}+c.c. \label{eq:gauss-ansatzqz}
\end{align}
The new key parameters in Eq.~\eqref{eq:gauss-ansatzqz} are the confocal parameter $b$, as well as the the beam waist radius $w_0$.
This more sophisticated solution therefore also justifies the criterion for using plane waves, namely that the interaction length $L$ should be much smaller than the Rayleigh range $z_R=b/2$ or in other words, the Rayleigh range should be very large, which corresponds to weak focusing, so that the amplitude is stable and only the exponential of the plane wave ansatz (p.w.) in Eq.~\eqref{eq:plane-waves-ansatz} determines the size of the amplitude $a$ in Eq.~\eqref{eq:gsol+ap}: 
\begin{equation}
    a\propto \int dz \frac{e^{i\Delta k z}}{\left(1+\frac{i2z}{b}\right)^{q-1}} \quad \overset{p.w.}{\Rightarrow} \quad \frac{L}{b}\ll 1 \; . \label{eq:gsol+ap}
\end{equation}
In Boyd's textbook\cite{Boyd92}, the resulting integral equation~\eqref{eq:intboydint} is solved for large crystals and focal positions far away from interfaces via the residue theorem:
\begin{align} \label{eq:intboydint}
    I(\Delta k, z_0,z)= \int_{-\infty}^{\infty} \frac{e^{i \Delta kz}}{\left(1+\frac{i2z}{b}\right)^{q-1}}dz\\
    \nonumber\\
    =\left\{ \begin{array}{cc} 0 &,\Delta k \leq 0 \;, \\ \frac{b}{2} \frac{2\pi}{(q-2)!}\left(\frac{b\Delta k}{2}\right)^{q-2} e^{-b\Delta k/2} &,\Delta k > 0 \; .\end{array}\right. \label{eq:intboydcon}
\end{align}
Eq.~\eqref{eq:intboydcon} shows that only for positive phase-mismatch, there is a signal when the focus is positioned in the crystal far away from the surfaces. Interestingly, even for full phase-matching, the signal in the bulk vanishes.
One has to note here, however, that the solution for the case $\Delta k =0$ is only valid for $q\ge3$, though. For SHG ($q=2$) and perfect phase-matching, the signal does actually not drop to zero. In the phase-matched case it holds $e^{i\Delta k}=1$. Therefore one can directly integrate the remaining $\frac{1}{\left(1+i2z/b\right)^{q-1}}$ term; for $q=2$, this leads not to a monomial but a logarithmic dependence, which does not vanish for large arguments. One can also picture the problem by using the analytic expression and trying to decrease $\Delta k$ to vanish. For THG and all higher harmonics, the transition is smooth, but not for SHG, where the result is $0$ or $b\pi$, depending on the order of $\Delta k\rightarrow 0$ and $q\rightarrow 2$. One can use the direct approach of the integral and successively decrease the symmetric boundaries to zero, whereas the phase term is isolated:
\begin{align}
    &\int_{-\infty}^{\infty}\frac{1}{1+i2z/b}dz=\left[\frac{ln(1+2iz/b)}{2i/b}\right]_{-\infty}^{\infty}\\
    &=\underset{z_2\to-\infty}{\lim_{z_1\to\infty}}\left[\frac{\ln(\sqrt{1+(2z/b)^2} e^{i \cdot arctan(2z/b)})}{2i/b}\right]_{z_2}^{z_1}\\
    &=\underset{z_2\to-\infty}{\lim_{z_1\to\infty}} \left[\ln \left(\frac{\sqrt{1+(2z_1/b)^2} e^{i \cdot \arctan(2z_1/b)}}{\sqrt{1+(2z_2/b)^2} e^{i \cdot \arctan(2z_2/b)}} \right) \frac{b}{2i}\right]\\
    &\overset{z_2=-z_1}{=}\lim_{z_1\to\infty} \frac{b}{2} \left[ \arctan{\left(\frac{2z_1}{b}\right)}-\arctan{\left(\frac{-2z_1}{b}\right)}\right]=\frac{\pi b}{2}\;.
\end{align}
For arbitrary boundaries, this difference of large terms is not a very suitable derivation, though.
However, one can also argue that the physical electric field is given by the real part, i.e. $2Re(E)=E+c.c$. Therefore, looking at the real and imaginary part of the electric field:
\begin{align}
    Re(E)&\propto\frac{1}{2}\int_{-\infty}^{\infty}dz\frac{1}{1+i2z/b}+\frac{1}{1-2iz/b}\\
    &=\frac{1}{2}\int_{-\infty}^{\infty}dz\frac{1-i2z/b+1+i2z/b}{1+(2z/b)^2}\\
    &=\frac{1}{2}\int_{-\infty}^{\infty}dz\frac{2}{1+(2z/b)^2}\\
    &=\frac{1}{2}\int_{-\infty}^{\infty}dx\frac{b}{1+x^2}\\
    &=\frac{1}{2}\left[b\cdot arctan(x)\right]_{-\infty}^{\infty}=\pi b/2 \;, \\
    Im(E)&\propto\frac{1}{2}\int_{-\infty}^{\infty}dz\frac{-4z/b}{1+(2z/b)^2}=0 \; .
    \label{eq:SHG-bulk-fin-signals-pm}
\end{align}
The real part is well defined for arbitrary boundaries and the imaginary part vanishes inside large crystals due to point symmetry. That means that SHG does not vanish in bulk if perfectly phase-matched and focused; however, the strength of the signal decreases with larger NA, as it depends linearly on the confocal parameter b.
To sum up, the analytical approximations in Eq.~\eqref{eq:intboydint} enable to include focusing into the description of the harmonic generation process. It is also shown, that focusing changes the value of the phase mismatch, which is for example also observed in recent experimental work\cite{ZRus21} of the authors. To obtain more realistic predictions, however, one also needs to utilize numerical strategies.\\

\section{Relations to Finite Elements Approach and available semi-analytical extensions}
Since the integrals of Eq.~\eqref{eq:intboydint} are strictly analytically calculable only for infinitely large crystals, they have to be solved numerically for finite boundaries, using finite numerical elements; one may, however, consider such a common numerically-aided approach which uses intuitive analytical trial expressions, including approximations, as semi-analytical. In contrast, 
more rigorous numerical toolkits solve the problem using the complete or only partially-approximated nonlinear wave-equation directly via summation over very basic structures. \\
Such an approach is employed for example when using plane waves in the angular spectrum representation, or when propagating an arbitrary polarization distribution via convolution with the point source propagator\cite{novotny2012principles,Sandkuijl2013,ZRus21,Spy2020LF,Spy2020FM}. Given enough computing power, they have the advantage, that they provide more accurate and sometimes more rich results, as they often solve the full 3D-problem without relying on symmetry. The latter enables to make very precise examinations of effects, which are not necessarily included when using specific trial solutions, like focus distortion and depolarization\cite{novotny2012principles,Spy2020LF,Spy2020FM,Hell93,NasseWoehl:10} for tight focusing conditions, as the trial solutions are mostly tied to a specific parameter regime. In these regimes, however, the semi-analytical approach, albeit being slightly less precise, has two other advantages. First, the needed computing power is much lower, which enables fast estimations and evaluations as well as scans over a larger parameter space of different setup conditions, such as focus depth, crystal thickness or focusing strengths, 
for example in order to optimize the setup. Secondly, apart from final results, i. e. the field distribution or output signal, single phenomenological parameters like the phase-mismatch, Gouy-phase, Rayleigh-range, etc. can be associated as beam parameters, and their influence can be understood more easily. This makes these approaches not only important for educational reasons, but also when extracting specific observables or improving the experimental method. It is critical, however, to keep track of the specific limits of the approach to make sensible interpretations. 
Such approaches do exist for many applications.\\
A similar approach can be found for example in the influential work\cite{BKl68} of Kleinman and Boyd, which discusses in detail the second-harmonic interaction of Gaussian beams and works out strategies to optimize the parametric output. There, the ratio of crystal length $l$ and confocal parameter $\xi=l/b$ is the key parameter, which has to be optimized. The in-depth discussion also uses a semi-analytical Gaussian approach where discussions on limiting cases are possible and provides an intuitive understanding of the situation.\\
Our work here deals with related topics, as far as parametric interactions of Gaussian beams are concerned. However, the main focus lies in the computation and understanding of parametric processes for material characterization, especially thin films and layered materials, using e.g. scanning microscopy. Here, varying thicknesses, substrates and also different parametric processes including SHG, THG, or CARS play an important role. The task is not necessarily the output optimization but to understand the influence that the optical setup and sample geometry have on the output signal. In order to be able to make reliable statements on the material properties, this illustrative disentanglement of the interplay of optics and material on the one side and material properties on the other side can be very useful. 
\section{\label{sec:Methods}Extending the Phenomenological Approach}
\subsection{Modeling of finite crystals and thin films}\label{sec:shg-tf-ln}
In order to describe nonlinear optical microscopy signals generated in bulk material or thin films more realistically, i.e. with finite thicknesses, multiple layers, and significantly tight focusing, the ansatz of the modeling presented in the last section has to be extended.\\
In the following, we will use a setup, where a transparent material, with good air/material transmission is bonded to a (reflecting) high index substrate. This conditions can be found for example in a lithium niobate slab on silicon, a typical layer structure which is often experimentally examined\cite{ZRus21} and typical in modern (quantum) optics\cite{MORU19,Liu22,LNOIRev22}, or when investigating 2D materials\cite{LZWX20}, which is depicted in Fig.~\ref{fig:setup-ill-tot}(a).
The basic ansatz of the used paraxial solution is taken from Boyd \cite{Boyd92}. Here, an incoming Gaussian fundamental beam is assumed and casted into a compact form as shown in Eq.~\eqref{eq:gaussbeam}:
\begin{equation}
    A_f(r,z)=\frac{a_1}{1+i2z/b} \cdot \exp\left(-\frac{r^2}{w_0^2(1+i2z/b)}\right)\;. \label{eq:gaussbeam}
\end{equation}
Here $a_1$ is a complex amplitude, $b=2z_R$ is the confocal parameter which corresponds to twice the Rayleigh-range, $w_0$ is the beam waist radius, and $r$ and $z$ are the cylindrical coordinates, where the beam propagates in $z$-direction and no angular dependence is present.\\
The well known coupled wave equation for nonlinear phenomena, including the slowly varying envelope approximation (SVEA), i.e. $\frac{\partial^2A_q}{\partial z^2}\ll k_q \frac{\partial A_q}{\partial z}$, will be used in cylindrical coordinates, cf. Eq.~\eqref{eq:coupledwaveeq2ii}.
\begin{equation}
    2ik_q \frac{\partial A_q}{\partial z}+\nabla_T^2A_q=-\frac{\omega_q^2}{\epsilon_0 c^2} P_q e^{(i\Delta k z)} \label{eq:coupledwaveeq2ii}\;,
\end{equation}
where $P_q$ is the amplitude of the contribution of the nonlinear material polarization, which acts as a source for waves with a frequency of $\omega_q=q\omega$.\\
Here, for $q^{th}$-harmonic generation (QHG),
\begin{equation}
\Delta k=qk_{\omega}n_{\omega}-k_{q\omega}n_{q\omega}    
\end{equation}
is the phase mismatch, where $q$ represents the order of the harmonic process, e.g. $q=2$ for SHG and $q=3$ for THG. Note that the refractive index $n=n(\lambda)$ is a quantity which depends on the wavelength $\lambda$ or frequency $\omega$ of the light. Thus, as in almost every material, the refractive index is not constant, a certain phase mismatch is almost always present. Usually, one also defines the distance $l_{c,class}$ up to which the waves are partially positively interfering and the harmonic signal is growing in the plane wave picture, cf. Eq.~\eqref{eq:lcoherence}, where for plane waves, the $\Delta k$ is the only contribution to the signal oscillations, as:
\begin{equation}
    l_{c,class}=\frac{\pi}{\Delta k}=\frac{\lambda_{\omega}}{2q} \frac{1}{n_{\omega}-n_{q\omega}} \label{eq:lcoherence} \text{ with }\left( \begin{array}{c} q=2 \text{, SHG}\\ q=3 \text{, THG} \\ ... \end{array}\right)\;.
\end{equation}
This equation can finally be compared to the extended ones of Eq.~\eqref{eq:exact-phi} and Eq.~\eqref{eq:approxnewcohl}.
It remains to be seen how the \mbox{experimental} conditions generate deviations from this plane wave case; however, it gives a rough estimation of the values to be expected.
In order to solve the differential equation, we use a trial solution which is similar to a Gaussian beam, but includes also a z-dependent amplitude, cf. Eq.~\eqref{eq:trialsolution}:

\begin{equation}
A_q(r,z)=\frac{a_q(z)}{1+i2z/b}\cdot \exp\left(-\frac{qr^2}{w_0^2(1+i2z/b)}\right)\;.
    \label{eq:trialsolution}
\end{equation}
By inserting Eq.~\eqref{eq:trialsolution} into the differential equation Eq.~\eqref{eq:coupledwaveeq2ii} one obtains in good approximation\cite{Boyd92} a solution for the amplitude in form of an ordinary differential equation\cite{Boyd92}. That can be integrated and yields Eq.~\eqref{eq:solutioncwe}:
\begin{equation}
    a_q(z;z_0)= c \int_{z_0}^{z}\frac{a_1^q \cdot e^{i\Delta k z'}}{(1+i2z'/b)^{q-1}} dz' \;. \label{eq:solutioncwe}
\end{equation}
Here, $c=\frac{iq\omega}{2nc}\chi^{(q)}$, where $\chi^{(q)}$ is the $q^{th}$-order nonlinear optical susceptibility, and $z$, $z_0$ are the start- and endpoint values of the material (and therefore the interaction region) determined \mbox{relative} to the focal position, which is zero in this parametrization. In the work of Boyd \cite{Boyd92}, presented in Sec.~\ref{sec:tbres}, the integral is solved analytically via extending the borders to infinity in an analysis of thick crystals. It shows the surprising result that even phase-matched THG does not deliver any harmonic signal, but needs to have a little positive phase mismatch in order to compensate for the additional Gouy-phase.
\\
In the case of thin films, however, as depicted in Fig.~\ref{fig:setup-ill-tot}(a), such a continuation of the integral is not possible; in fact, even typical scenarios of surface-near scans [cf. Fig.~\ref{fig:setup-ill-tot}(b)] are not described by that approximation. It has to be solved numerically for finite boundaries. Furthermore, especially when samples are mounted on a strongly reflecting substrate like silicon, one has to consider the effects of reflection and transmission as well.
Since in a typcial case of lithium niobate (LN), the interface between LN and air has a high transmission \mbox{$T_{\perp}(\lambda \approx 0.85...0.9~\mu\text{m})\sim 85\%$}, generally, it is sufficient to take into account only a small number $n$ of reflections; especially for the fundamental beam, interferences of the multiple reflections change the basic properties of our assumptions as they are related to the harmonic in a nonlinear fashion and have to be taken into account in detail; the issue here is that the polarization in the NLO material slab of thickness $d$ would be a large sum of positive direction propagating waves $z\rightarrow(z+2dn)$ and  negative direction propagating waves 
$z\rightarrow (2nd-z)$.
This looks like the following Eq.~\eqref{eq:prop-waves-exp}, modulo the transverse parts:
\begin{align}
P_{NL}\propto \left[\frac{e^{ikz}}{1+i2z/b}+\sum_{n=1} \left( \frac{c_{neg,n}e^{ik(2nd-z)}}{1+i2[2nd-z]/b}\right.\right. \label{eq:prop-waves-exp} \\
\left.\left. +\frac{c_{pos,n}e^{ik(z+2nd)}}{1+i2[z+2nd]/b}\right)+c.c.\right]^{q} \;. \nonumber
\end{align}
Here, $c_{neg,n}$ and $c_{pos,n}$ are the amplitude coefficients for propagation in positive direction and negative direction, respectively, which are reflected $n$ times; they consist of the Fresnel coefficients for the corresponding reflections (and damping factors).
When searching for the $q^{th}$ harmonic, we need the $e^{iqkz}$ and the complex conjugate term. Therefore, we can treat the counter- and co-propagating waves independently and both sums can be simplified: We can group all elements of a respective sum together in a geometric series, when approximating the denominator in a damping factor $A_{damp}$; each series is dominated by a constant phase factor of $e^{ik\cdot2d}$ and the corresponding Fresnel-coefficients then.\\
Thus, one can calculate the two contributions with two separate integrals for forward and backward propagating fundamental, where the generated harmonics may interfere afterwards. The corresponding thin film interference is included via 
multiplying the integrand with a geometric series to the power of $q$, which depends on the thickness $d$ of the slab. 
This dependence can also be implemented for the harmonic light, which shows only the linear interference which can be as well modulated with a \mbox{geometric} series as shown in Eq.~\eqref{eq:geoseries}.
\begin{align}
    \sum_{0}^{\infty}q^n=\frac{1}{1-q} \; \text{, for }q<1 \; , \label{eq:geoseries}\\
    \Rightarrow q=c_{x} \cdot e^{i2k_xd}\; ,
\end{align}
where $k_x$ is the fundamental or harmonic k-vector, and $c_x(\omega)=r_{LN,Si}(\omega)r_{LN,Si}(\omega)$ summarizes the Fresnel-coefficients corresponding to the reflections at the interfaces for the respective frequency $\omega$, i.e. the fundamental or harmonic frequency. In addition, depending on the point of observation, the integrals obtain further Fresnel-coefficients, e.g because the signal is detected in reflection geometry, the positive direction contribution gets a global $r_{LN,Si}(\omega_q)$, whereas the integral of the negative direction propagating waves naturally shows $r_{LN,Si}(\omega)$.
In general, further damping coefficients can be introduced as well. The series produces a characteristic pattern of oscillations and  for the fundamental wave it holds that the larger $q$, the more significant the oscillation gets, as $q$ sits in the exponent.\\
One has to  note, that obviously the beam, when propagating, diverges with a polynomial not an exponential dependence. But as long as the transmission out of the material is quite large, the thin film interference can be very roughly approximated via a geometric series using a damping factor as mentioned above, as the damping by the transmission "losses" dominates. In comparison to the high transmission losses, the amplitude change due to the propagation, which evolves with $\frac{1}{(1+i2z/b)^{q-1}}$, does not result in a significant decrease for crystal thicknesses of order 
$t \leq 2b$.
Furthermore, when $L\gg b$ the additional contribution due to reflected light is actually also getting increasingly weaker as the intensity far away from the focus also generates less signal contribution, so that the ansatz remains viable.
For different samples, especially with significantly smaller transmission coefficients between air and material, it might be necessary, however, to use the exact expression for the material polarization, as shown in Eq.~\eqref{eq:prop-waves-exp}.
\\
In fact, the geometric series obviously contributes oscillations much faster than that of the phase mismatch; they are correlated to the thin film interference of the fundamental and harmonic light, and clearly observable in Fourier-transforms\cite{ZRus21,ZRus22} of simulation and experiment, but concerning the coherence length, are more of cosmetic nature.\\
It should be noted here, that for SHG, the calculation is in that sense simpler as only the NLO material, in this case the lithium niobate slab, contributes to the SHG signal due to the vanishing $\chi^{(2)}$ in all other materials (air, silicon) at hand. Therefore the aforementioned parameter $c$ of Eq.~\eqref{eq:solutioncwe} is not too important, and we end up with an according integration of the first incoming and (multi-)reflected beam contributing to a signal which is further modulated by reflections of the higher harmonics.\\
For THG, one needs to take into account the signal produced in air, or any other cladding material, and finally sum over both, so that a more complex interference pattern is expected. This can be described by a sum of integrals, as sketched in Eq.~\eqref{eq:int-change}:
\begin{eqnarray}
&c\int_{z_0}^{z_{max}}dz \rightarrow c_{air}\int_{-\infty}^{z_0=z_{LN}}dz+c_{LN}\int_{z_{LN}}^{z_{max}}dz \; . \label{eq:int-change}
\end{eqnarray}
Note, that in such a case of contributions from different layers, the propagation to a common endpoint needs to be taken into account. Here, the air signal can be reflected, i.e. multiplied with the reflection coefficient, or enter the crystal, i.e. multiplied with the same geometric series, the harmonic signal generated in the material encounters.\\
In this scenario, the exact ratio of the nonlinear optical susceptibility of air $\chi^{(3)}_{air}$ and lithium niobate $\chi^{(3)}_{LN}$ is of critical importance, as they define the strength and phase of the respective contribution, as will be shown in the result section. 
\subsection{The phase mismatch and coherent interaction length in Gaussian beams}
In fact, the changes in the phase evolution that stem from focusing, should be made quantifiable. As seen in different publications\cite{ZRus21,Boyd92}, the focusing conditions influence the observed coherent interaction length and the overall output signal. In some cases\cite{Boyd92} focusing compensates for the positive phase mismatch, whereas it may also decrease the coherent interaction length, as observed for negative phase mismatch\cite{ZRus21}.
In order to get a better understanding of this phenomenon one can reformulate the integrand of Eq.~\eqref{eq:solutioncwe} to the exponential form:
\begin{align}
a(z)&=\int \frac{e^{i\Delta kz'}}{(1+i 2z'/b)^{q-1}}dz'\\
& \Rightarrow \frac{e^{i \Delta kz}}{(1+i 2z/b)^{q-1}} = e^{i\Delta kz}\left[\frac{1-i2z/b}{1+4z^2/b^2}\right]^{q-1}\\
&=e^{i\Delta kz}\frac{e^{i\cdot(q-1)\cdot arctan(-2z/b)}}{(\sqrt{1+4z^2/b^2})^{q-1}} \;.
\end{align}
Extracting the phase leads to Eq.~\eqref{eq:com-phase}, as:
\begin{align}
    \phi(z)&= \Delta k z +(q-1)\cdot \overbrace{arctan(-2z/b)}^\text{Gouy-Phase} \label{eq:com-phase}\\
    &\approx \Delta k z -(q-1)\cdot 2z/b \; .
\end{align}
Demanding that the coherence length $l_c$ is the length up to which wavelets are added, i.e. $\Delta \Phi(z)\overset{!}{=}\pi$, we obtain Eq.~\eqref{eq:exact-phi}:
\begin{align}
    &\Delta \Phi=\pi=\phi(x_1)-\phi(x_0)\;, \quad \text{where} \;\; l_c=x_1-x_0 \label{eq:exact-phi} \\
    &\text{with  } x_0=0, \;\;  L\ll b \;\;\; \Rightarrow  l_c\approx\left|\frac{\pi}{\Delta k -(q-1)\cdot2/b}\right| \;. \label{eq:approxnewcohl}
\end{align}
Here, the approximation of Eq.~\eqref{eq:approxnewcohl} gives a good first estimation and is valid for $|2z/b|< 1$, meaning the Rayleigh range is larger than the interaction region. That is the regime, where the $arctan(2z/b)$ is almost linear.
The regime of Eq.~\eqref{eq:approxnewcohl} and the transition to Eq.~\eqref{eq:exact-phi} can be particularly well prepared and examined, when one is able to control the thickness of the sample, as in examinations of thin films/wedges, as is shown in the following sections. Furthermore, one can observe, that for high phase mismatches, meaning that $\Delta k\gg (q-1)2/b$, the Gouy-phase contributes only minor corrections, so that the significance of the Gouy-phase is also dependent on the wavelength and the order $q$ of the process.\\
Note, that when contemplating increasing crystal thicknesses or tighter focusing conditions, i.e. $L\ge b/2$ the approximation is not valid and one better uses the full $arctan(2z/b)$-term. In the following, we will also use the complete transcendental equation and apply a common root finding algorithm, i.e. Newton's method, to find an exact solution.\\
In fact, both formulas imply, that on the one hand a positive phase mismatch $\Delta k >0 $, i.e. the missing quasi-momentum, and the Gouy-Phase can partially compensate each other, so that the total effective phase mismatch in the relevant focal region can remain rather small, as is also observed in Boyd's textbook\cite{Boyd92}. On the other hand, for $\Delta k < 0$ the total phase mismatch is always increased, so that the coherence length for negative phase mismatch gets reduced for all wavelengths\cite{ZRus21}.
\par
However, the exact phase evolution implies more modifications in comparison to the case without focusing. Therefore, in summary and in order to have a common interpretation of the nomenclature, one has to carefully differentiate the observables, which are commonly associated to the term "coherent interaction length", or colloquially shortened to "coherence length". The coherent interaction length $l_c$ is the interaction distance up to which the generated wavelets interfere constructively. This corresponds to the distance of a minimum of local NLO intensity and the following maximum. Experimentally, as it is easier to read off, one often uses the distance from minimum to minimum\cite{ZRus21}. Thus, the coherence length therefore contains a total phase shift of the wavelets of $\pi$ and for the experimentally determined minimum-to-minimum distance it is $2\pi$, as visualized in Fig.~\ref{fig:SHG-thickness-lambda-foc}(a). One could call all these different distances in relation to their phase as $l_{\pi}$, $l_{2\pi}$. For the plane wave case, which is often implicitly assumed, as the phase $\phi(z)=\Delta k z$ changes linearly, it naturally holds that $2\cdot l_{\pi}= l_{2\pi}$, so that both are directly linked. Moreover, \textbf{every single oscillation} in the plane wave case is the same; it does not matter if one looks at the first minimum-to-minimum distance or any other. By definition, this behavior also holds true for the linear approximation of Eq.~\eqref{eq:approxnewcohl}. But as was already shown in Eq.~\eqref{eq:exact-phi}, for focused systems, the exact phase evolution is no longer linear. Here, due to the $arctan(2z/b)$-term, $2\cdot l_{\pi}\neq l_{2\pi}$; for example, having a negative $\Delta k$, the $-arctan(2z/b)$ term adds additional phase, so that $\pi$ is reached faster than in the plane wave case. For larger $z$, the $arctan(2z/b)$ converges to a constant, however, so that the following distances are larger, i.e. $2\cdot l_{\pi}<l_{2\pi}$. This relation then holds for all subsequent oscillations, until the $arctan(2z/b)$ is almost constant and the oscillation converges to the plane wave case. From the nonlinearity it also follows that $l_{\Delta \phi}$ also depends on the position of the focus relative to the optical material. When imagining the slab far away from the focus, the extra phase from the $arctan(2z/b)$ has no effects, but placing it within the Rayleigh-range it does, so that even the placement of the focus, e.g. on the interface or at $z=z_R$ produces different results.\\
To clarify which observable is dealt with, we will use $l_{\Delta \phi}$, e.g. $l_{2\pi}$ for the first minimum-to-minimum distance. Furthermore, the entry point of the beam into the crystal will be given as a reference point, which will in general be the surface at $z_0=0$.

\section{Case Study}
\begin{figure}
    \centering
\includegraphics[width=0.71\linewidth]{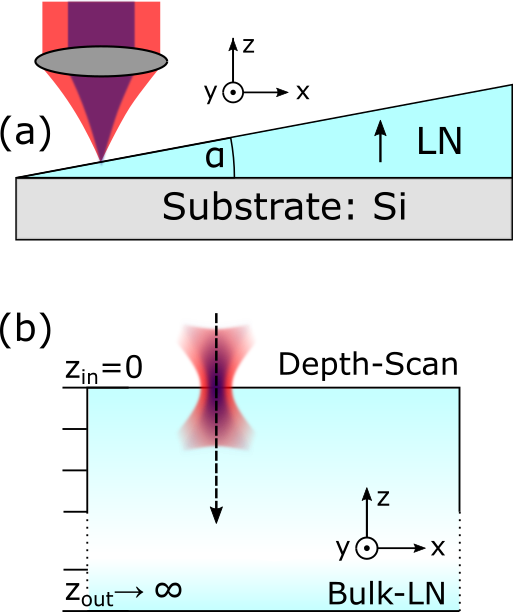}
    \caption{Illustration of \textbf{(a)} the z-cut LN wedge sample used in the case study to examine the thickness dependence of the nonlinear signal for thin films. The angle $\alpha$ is exaggerated to improve visibility. Subfigure \textbf{(b)} visualizes the geometry of a depth-scan into a semi-infinite crystal, which is used as the basic geometry to record the data displayed in Fig.~\ref{fig:surface-shg-ges} and Fig.~\ref{fig:mismatch-bulk-cases}. Note that $z_f=0$ corresponds to a focus position at the top interface and positive $z_f$-values correspond to a focus position $z_f$ inside the crystal. From the viewpoint of the Gaussian beam of Eq.~\eqref{eq:gaussbeam}, which is centered at the origin, the integral boundaries (or the integration variable) have to be adapted for the actual calculation, e.g. $z'_{in}=-z_f$ as the adapted lower boundary.}
    \label{fig:setup-ill-tot}
\end{figure}

In the following subsections, we study different NLO processes taking place in the model system of thin film LN (TFLN) on silicon, and aim at working out the influence of different parameters on the final signals. This thin film system is schematically depicted in Fig.~\ref{fig:setup-ill-tot}(a), where we have used a wedged z-cut LN sample for variable thickness testing.
All following studies will use this system, except the first, which deals with SHG in bulk LN. The values of $n_{LN}(\lambda)$ are calculated via the Sellmeier equation, using the coefficients\cite{Zelmon:97} as tabulated in the work of Zelmon et al.


\subsection{SHG in bulk LN}\label{sec:shg-in-bulk}
\begin{figure}
    \centering
    \includegraphics[width=0.99\linewidth]{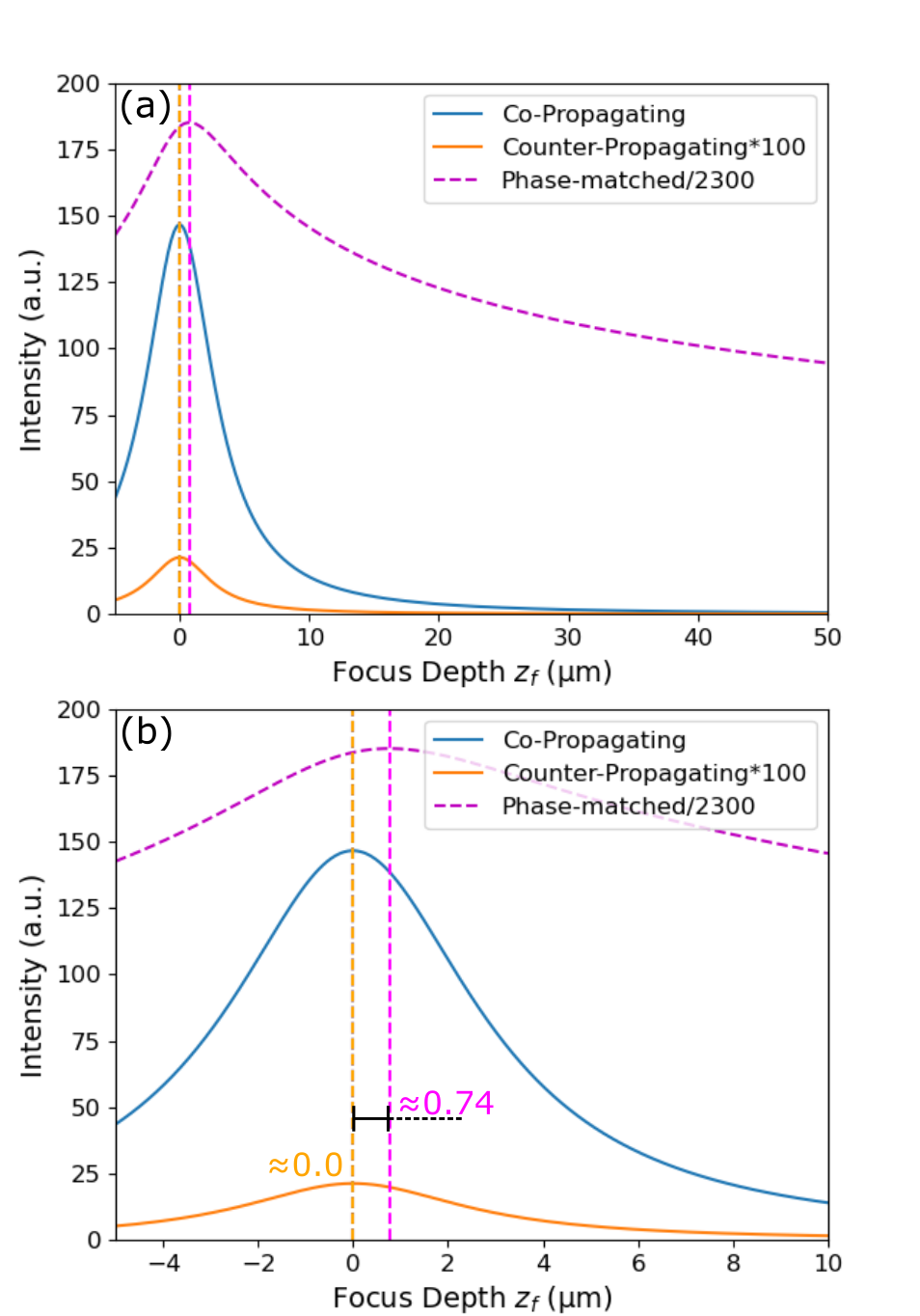}
    \caption{Results for surface-near second-harmonic depth scans [cf. Fig.~\ref{fig:setup-ill-tot}(b)] for co- and counter-propagating signal in a semi-infinite LN crystal. The fundamental wavelength of 850~nm and a NA of 0.45 are employed for this simulation, alongside an actual crystal thickness of $t=3$~mm. One observes the experimentally well known signal peak at the surface, typical to large phase mismatch. A hypothetical phase-matched case is also given for discussion; \textbf{(a)} shows an overview down to $50$~$\mu$m into the crystal, whereas \textbf{(b)} shows a close-up of the interface region. Here, with this moderate focusing, one observes that the phase-matched case shows the peak to be well inside the material, while the other cases have their peak approximately at the surface.}
    \label{fig:surface-shg-ges}
\end{figure}

Before promoting LN to a thin film, we want to take a look at a LN bulk crystal once more. As already mentioned, the integral of the amplitude has to be solved numerically, in general. Even for large crystals, in the case of surface-near scans, the setup does not fulfill the condition of Eq.~\eqref{eq:intboydcon} allowing for infinite integration boundaries. These scans, however, are used for example for ferroelectric domain mapping \cite{Spy2020FM}. As the crystal is thick, the upper boundary can be set to infinity, whereas the entering point $z_0$ is near zero. The experimental situation is depicted in Fig.~\ref{fig:setup-ill-tot}(b).
The corresponding amplitude can be computed with Eq.~\eqref{eq:bulkcase}:
\begin{align}
a_q= c_{LN} \int_{z_0}^{\infty}\frac{a_1^q \cdot e^{i\Delta k z'}}{(1+i2z/b)^{q-1}}dz' \;. \label{eq:bulkcase}
\end{align}
Now,
we can in principle observe co- and counter-propagating second-harmonic signal, where the latter has a very large, but positive phase-mismatch $\Delta k =2|k_1|+|k_2|$.
As the crystal is very thick and bounded by an LN/air-interface on both sides, multiple reflections can be neglected due to high interface transmission, scattering and damping of the signal.\\ 
In Fig.~\ref{fig:surface-shg-ges}(a) one can observe the typical significant surface-near SHG\cite{Spy2017II,Spy2020FM}. As can be extracted by the bulk integrals of Eq.~\eqref{eq:intboydcon} where the signal vanishes in an infinite bulk crystal, one sees SHG at the surface despite the negative (co-propagation) or very large (counter-propagation) positive mismatch. The counter-propagating signal is more than two orders of magnitudes weaker, though, simply due to the much shorter coherence length. As a side note, this also means, that in a back-reflection setup, depending on the NA and the actual crystal thickness and therefore collection efficiency of the diverging signal, it is possible to observe mainly reflected co-propagating light, as the transmission losses at the backside are roughly "only" $85\%$. Here, the co-propagating light-signal would still be more than one order of magnitude larger, provided it can be collected efficiently.

     
     %

Fig.~\ref{fig:surface-shg-ges}(a) also shows a case of a phase matched \mbox{($\Delta k=0$)} scenario, which may be achieved in birefringent crystals by choosing specific wavelengths and scattering geometries. Since the  corresponding signal is much stronger, it is scaled down, to be comparable; one can state, that much more signal is generated, but the main peak still lies near the surface. However, it is not exactly centered at the surface. It validates the calculations of Eq.~\eqref{eq:SHG-bulk-fin-signals-pm} that there remains a (weak) signal from the bulk, but also demonstrates, how strong the influence of the Gouy-phase is, functioning as an effective phase-mismatch here, when compared to a plane wave, where the signal would monotonically grow with crystal thickness.
\par 
It is interesting to consider the influence of the NA and the coherence length at this point. The NA, i.e. strength of focusing, on the one hand is the dominant quantity for the focal length $b$  and for strong focusing this length dominates the effective interaction length and is also responsible for the rate, at which the Gouy-phase changes, as $\phi_{Gouy} \propto arctan(\frac{2z}{b})$. On the other hand, the phase mismatch $\Delta k$ determines the length scale on which constructive interference is possible. It is modified, however, by the Gouy-phase, as shown in Eq.~\eqref{eq:com-phase}. Therefore, the final results are determined by the ratio of $\Delta k$ to $\frac{2}{b}$.
\par
In this context, the dependence of the exact position of the signal peak needs further consideration. One would naively consider that to get the most signal, the focus of the pump beam should lie inside the material. 
In a closer examination, one can actually observe, however, that the maximum is mostly symmetrically distributed at the surface, cf. Fig.~\ref{fig:surface-shg-ges}(b). Although, the positions of these example cases, motivated by experimental work\cite{ZRus21}, only show tiny deviations, i.e. $\sim 10$~nm, from the surface, when choosing other wavelengths that produce smaller phase-mismatch values or even anomalous dispersion, the deviation can be significant, i. e. $\sim 1$~µm, and will be easily observable. Therefore, we will also examine scenarios with decreased mismatch in the following, where  $\Delta k$ is set to be within the same order of magnitude as the inverse of the focal length, i.e. $\frac{1}{b}\sim \Delta k$.
\par
For $\Delta k \gg \frac{2}{b}$, the signal concentrates at the surface $z_{foc}\approx 0$ and has a width within the same order of magnitude as the focal length, as also observed in experiments\cite{Spy2020FM}. The latter is due to the fact, that with high $\Delta k$, i.e. $l_c\ll b$, the rapid $\Delta k$-oscillations prohibit constructive interference on a larger length-scale. Therefore, whenever the intensity maximum enters the crystal, the strongest signal can be detected. When the rest of the beam enters, the signal decreases again due to corresponding destructive interference.\\
One may consider, that due to the Gouy phase, the coherence length of counter-propagating light is (slightly) increased in dependence of the NA, but the mismatch for counter-propagating light is in general so large, that the effect is almost negligible, i.e. the behaviour for co- and counter-propagation is similar in this case.
\par
When $\Delta k$ is significantly smaller, the deviation from the surface is modified by the interplay of the phase mismatch and the Gouy-phase, as both can contribute to the total phase evolution via $\Phi(z)= \Delta k-\arctan(2z/b)$.
One can start from two different limiting cases. 
\par

\begin{figure}
    \centering
    \includegraphics[clip,width=\columnwidth]{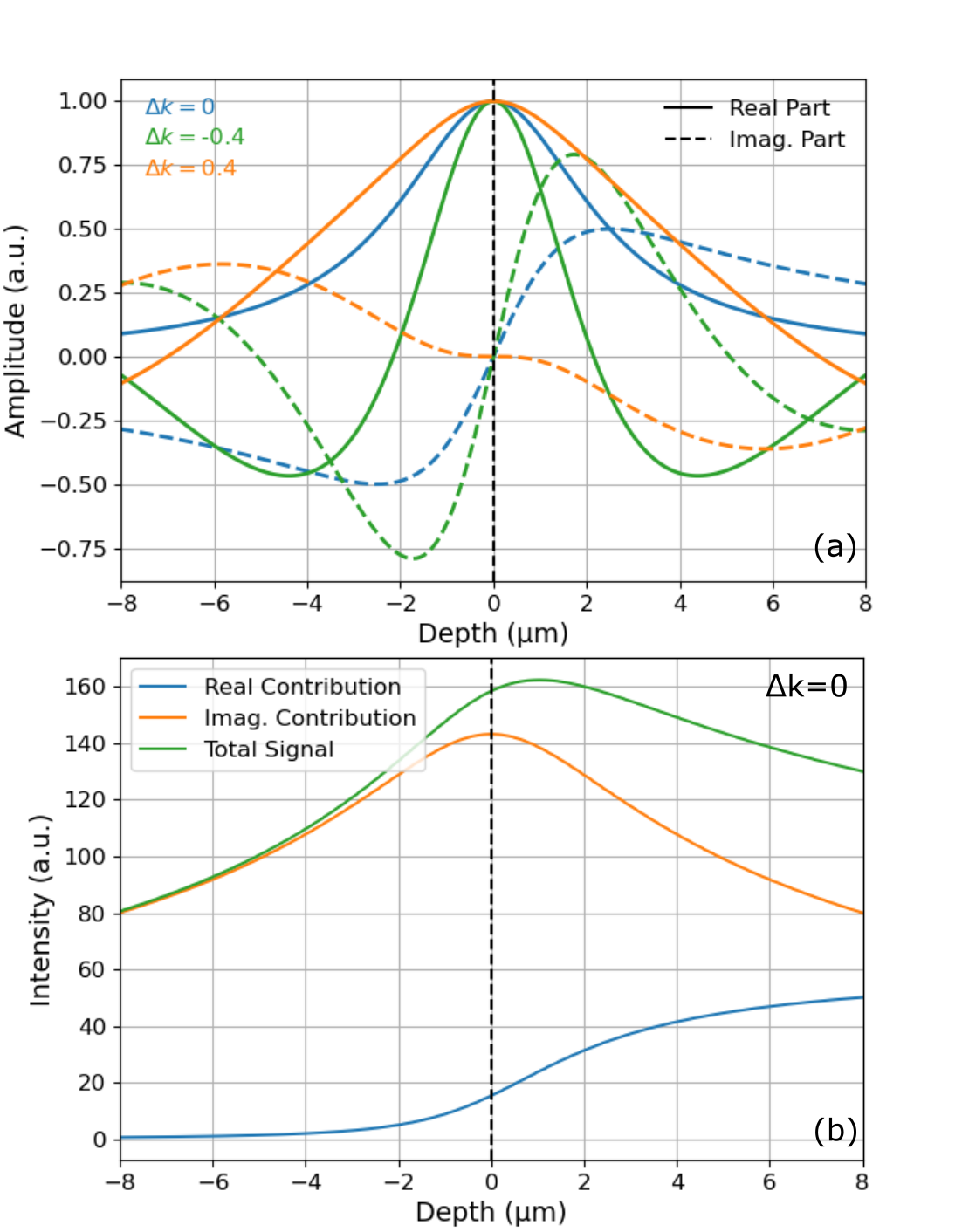}
    \caption{Example simulations with confocal parameter $b=5$ and crystal thickness of $t=0.5$~mm. Subfigure \textbf{(a)} shows the real and imaginary part of the integrand of Eq.~\eqref{eq:sig-int-parts} for different values of phase mismatch $\Delta k$. In subfigure \textbf{(b)}, you can see an example of the absolute squares of the computed integrals for $\Delta k=0$, where the contributions of real and imaginary part and the total signal are illustrated separately. }
    \label{fig:contributions}
\end{figure}

When the focusing is significantly stronger or the phase mismatch very small, i.e. $\Delta k \ll \frac{2}{b}$ the  Gouy-phase determines the evolution of the phase from $-\pi/2$ to $\pi/2$. It helps to look at the real and imaginary part of the complex integrand, as given in Eq.~\eqref{eq:sig-int-parts}:
\begin{align}
a_q&\sim \int_{z_0}^{\infty}\frac{e^{i(\Delta k x- \arctan(\frac{2x}{b}))}}{\sqrt{1+\left(\frac{2x}{b}\right)^2}}dx\label{eq:sig-int-parts}\\ 
&=\int_{z_0}^{\infty}\frac{\cos(\Delta kx-\arctan(\frac{2x}{b}))}{\sqrt{1+\left(\frac{2x}{b}\right)^2}}+i\frac{\sin(\Delta kx-\arctan(\frac{2x}{b}))}{\sqrt{1+\left(\frac{2x}{b}\right)^2}}dx \;. \nonumber
\end{align}
As is illustrated in Fig.~\ref{fig:contributions}(a), the real part, i.e. the $\cos[\arctan(2z/b)]$ part, is symmetric and  for $\Delta k = 0$ always larger than zero, so that the optimum is situated in the center of the crystal. The $\sin[\arctan(2z/b)]$ part is antisymmetric, changes sign at the origin and is largest when the focus is positioned at the surface. Then, the total optimum compromises both contributions and, as shown in an example calculation in Fig.~\ref{fig:contributions}(b) for the phase matched case, is near the surface but slightly inside the material for thick but finite crystals.
\par 
As a side note, the exact position depends slightly on the exact thickness via the cutoff as well, which has to be introduced to regularize the logarithmically divergent integral of the imaginary part. The physical interpretation of this cutoff is nothing else as the crystal thickness; Obviously, the thinner the crystal, the more the  damped oscillatory behavior of the integrand becomes visible in the position of the maximum in dependence of the crystal length, so that even maximum positions slightly above the surface are observable under certain conditions. In order to minimize these effects, the numerical analysis is carried out with a realistic bulk crystal thickness of $t=3$~mm, so that the amplitude of the oscillation of the integrand at the upper border of the integral is three orders of magnitudes smaller than the maximum.
\par
For scenarios in between, i.e. $2/b \sim \Delta k$, the relation between $\Delta k$ and $b$ is crucial. When $\Delta k < 0$, the phase-mismatch and the Gouy-phase add and the oscillating contributions lead to an overall weaker signal. Furthermore, for increasing phase mismatch, due to the introduced oscillations, cf. Fig.~\ref{fig:contributions}(a), the maximum of the integral of the $\cos[\Delta kz+\arctan(2z/b)]$ moves from deep into the crystal in the direction of the surface, avoiding the negative contribution left of the position of the first root, which appears on the left of the central maximum. The total maximum therefore moves continously to the surface; it may even move slightly above the surface, when the total negative contribution of the symmetric part is larger, which depends on the cutoff, i.e. the crystal thickness. Then, with a further growing phase mismatch, the maximum position will oscillate around the surface and rapidly converge to the surface position for large $\Delta k$ reaching the case $\Delta k \gg \frac{2}{b}$, where the phase-mismatch dominates, cf. Fig.~\ref{fig:surface-shg-ges}(b).
\par
When $2/b \sim \Delta k$ but $\Delta k > 0$, the phase mismatch and the Gouy-phase can partially compensate. This leads to a very flat evolution of both integrands, as is demonstrated in Fig.~\ref{fig:contributions}(a). Although the maximum of the symmetric part moves slowly to the surface, the antisymmetric contribution almost vanishes in the focal region and the total maximum moves closer to the maximum position of the symmetric part. For larger $\Delta k$, the evolution gets sinusoidal again and converges in the direction of the surface again as in Fig.~\ref{fig:surface-shg-ges}(b).
\par
\begin{figure}
    \centering
    \includegraphics[width=0.99\linewidth]{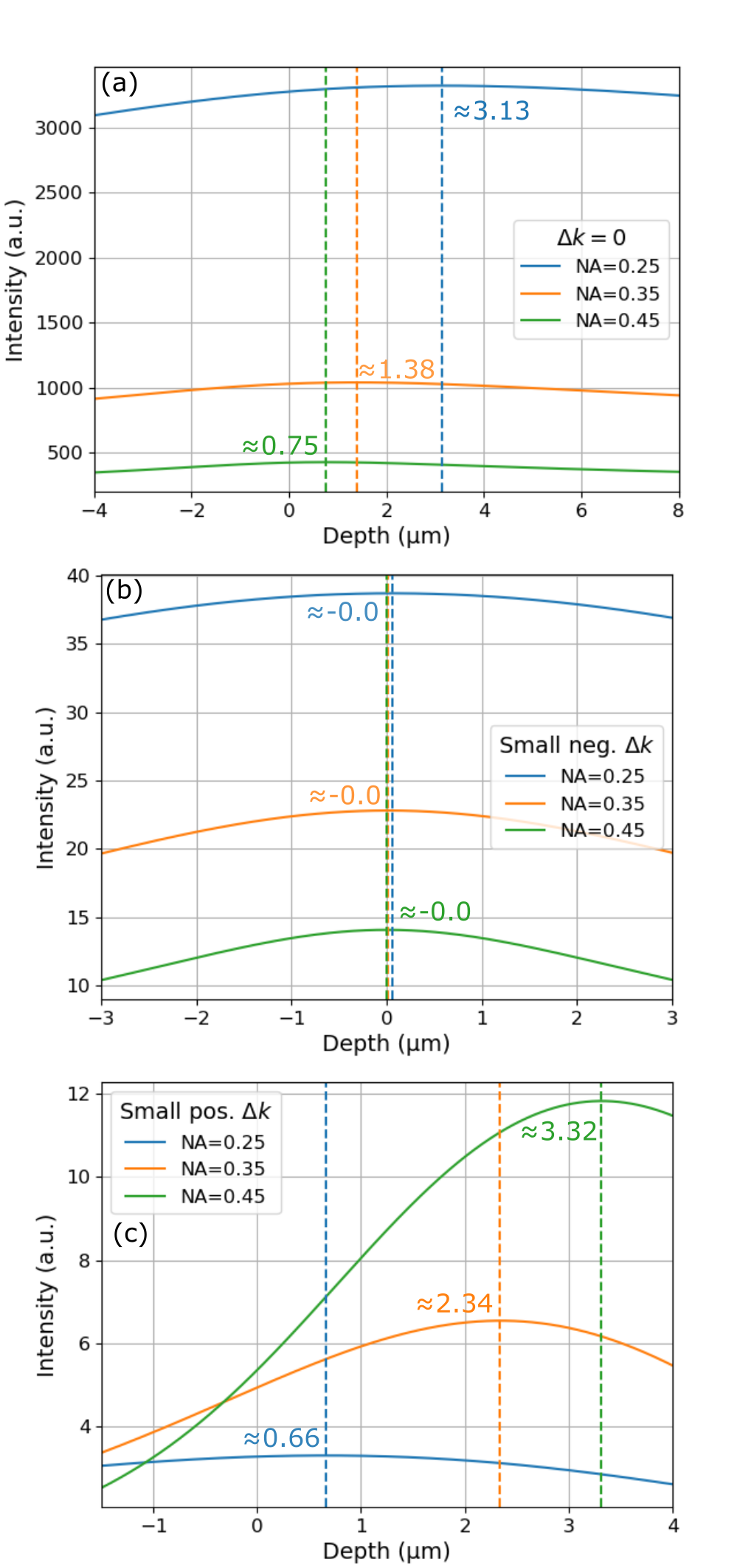}
    \caption{Surface-near SHG depth-scans [(cf. Fig.\ref{fig:setup-ill-tot}(b)] for different values of $\Delta k$ and different NAs in a semi-infinite LN crystal. The fundamental wavelength of 850~nm and a NA of 0.45 is employed for the simulation and the original $\Delta k$ values for co- ($\Delta k_{co}$) and counter-propagation ($\Delta k_{count}$) at 850~nm are changed to $\Delta k'$, to illustrate possible different signal behavior. One observes that \textbf{(a)} for $\Delta k'=0$, the maximum moves from inside to the surface with increasing NA; \textbf{(b)} for small negative mismatch ($\Delta k' \rightarrow 0.04 \Delta k_{co}$), the maximum is slightly inside or outside the surface, to evade the additional mismatch due to the Gouy-phase, and moves to the surface for increased focusing; \textbf{(c)} for small positive $\Delta k$ ($\Delta k \rightarrow 0.01 \Delta k_{count}$), the maximum also lies inside the crystal and moves inward for higher NAs as the Gouy-phase compensates the mismatch until its gets too large and can be approximated with case (a).}
    \label{fig:mismatch-bulk-cases}
\end{figure}

Finally, if in the former medium regime cases the NA is increased, the dominance of the Gouy-phase grows again. Here, all optimum positions shift increasingly to the surface, cf. Fig.~\ref{fig:mismatch-bulk-cases}(a) and
Fig.~\ref{fig:mismatch-bulk-cases}(b), as $\frac{2}{b}$ increases while also the absolute scale $b$ decreases; 
the only exception is the $\Delta k >0$ case, where the Gouy-phase is still compensating for the positive phase mismatch.\\
In conclusion, this means that the position of the surface does not necessarily coincide with the position of strongest intensity, at least not when the phase mismatch is small or of similar magnitude compared to $\frac{2}{b}$ and especially when it is positive, i.e. the chosen wavelength range shows anomalous dispersion. This may be relevant when accurate positions in depth need to be determined via SHG microscopy.

     %

\subsection{SHG in TFLN}
Next, we will consider SHG in the thin LN slab on silicon as shown in Fig.~\ref{fig:setup-ill-tot}(a). Here, two components can be expected. We can have co-propagating and counter-propagating second-harmonic light, where the former plays the dominant role. We describe scans along the direction of growing crystal thickness; a probable experimental situation is depicted in Fig.~\ref{fig:SHG-thickness-lambda-foc}(c). 
The scans normally are considered with the focus lying at the surface $z_{in}=0$, but can also be described for a variable focus depth. The effects of such differing focal depths are dealt with in the second part of this subsection.\\
\subsubsection{Signal evolution for fixed focus position}
We now employ the discussed approach building on Eq.~\eqref{eq:solutioncwe} with finite boundaries. We also take into account multiple reflections in the slab of thickness $L$ by including the nonlinear polarization generated by subsequent reflections, as shown in Eq.~\eqref{eq:prop-waves-exp}, as well as in the modulation of the fundamental and harmonic signal via the geometric series, as shown in Eq.~\eqref{eq:geoseries},  where the series belonging to the fundamental is set to the power of $q$:
\begin{align}
    S&(L,z_f=0;\Delta k)\propto \nonumber\\
    &\left[\int_0^L r_{LN,Si}(\omega_q)\frac{a_1^q e^{i\Delta k z'}}{(1+i2z'/b)^{q-1}} dz'\right. \nonumber \\&\left. + \int_0^L r_{LN,Si}^q(\omega)\frac{a_1^q e^{i\Delta k(z'+L)}}{(1+i2(z'+L)/b)^{q-1}}dz'\right]  \nonumber \\ 
    &\cdot \left(\frac{1}{1-c_x(\omega) e^{i2kL}}\right)^q\left(\frac{1}{1-c_x(\omega_q)e^{i2k_qL}}\right)\;.
\end{align}

\begin{figure}
    \centering
    \includegraphics[width=0.99\linewidth]{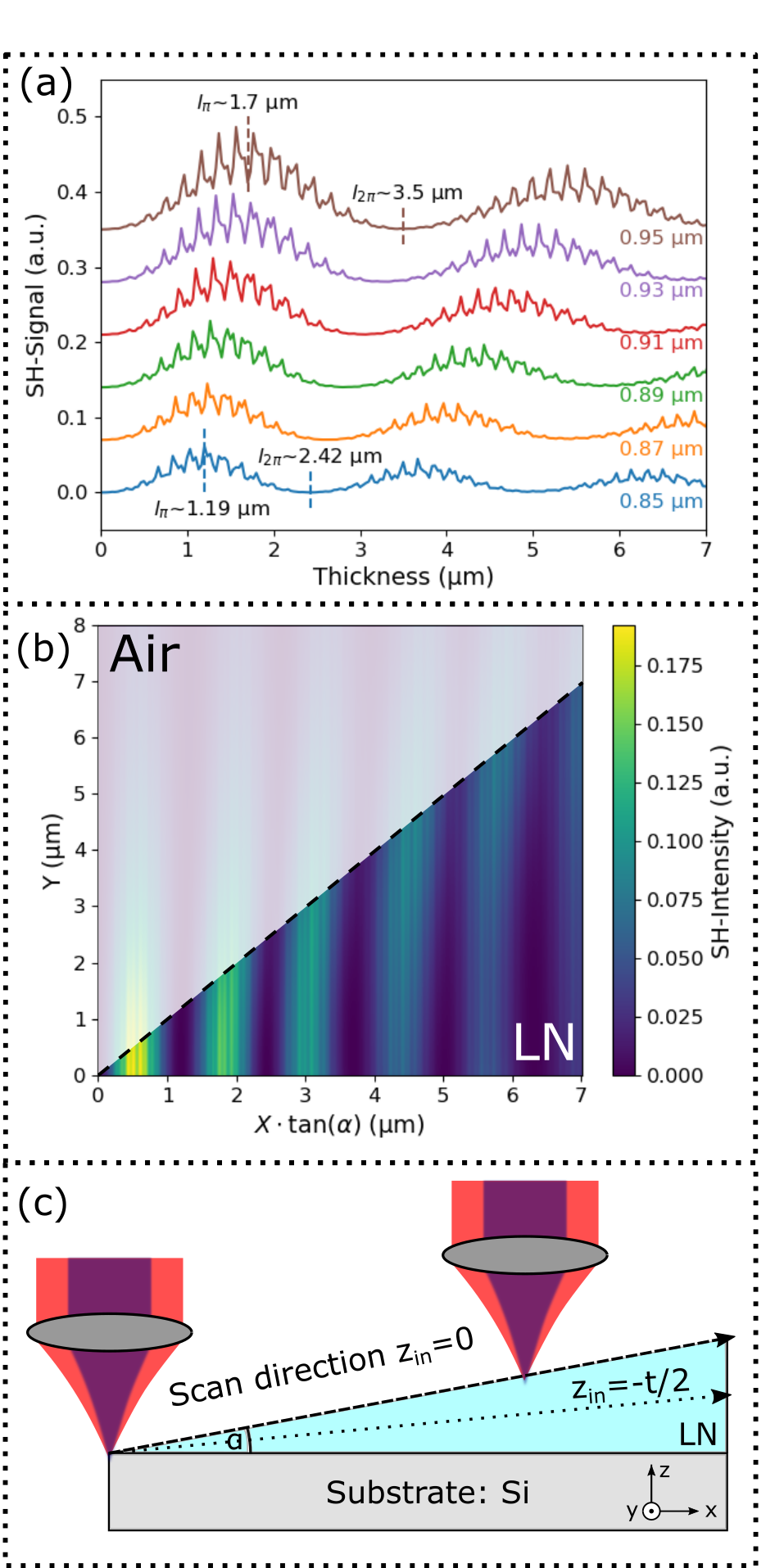}
    \caption{SHG signal of \textbf{(a)} a line-scan  for different LN thicknesses with $z_{f}=0$. The wavelength is varied from 850~nm to 950~nm and the NA is 0.45. One observes that the coherence length is influenced similar to experiments \cite{ZRus21}. In \textbf{(b)} the corresponding cross-section of the wedge is shown for $\lambda=850$~nm. In the depicted depths the evolution changes only marginally. In \textbf{(c)} the measurement setup is sketched, i.e. for (a) a line-scan on the surface was carried out.}
    \label{fig:SHG-thickness-lambda-foc}
\end{figure}

One observes in Fig.~\ref{fig:SHG-thickness-lambda-foc}(a) a similar behavior as in the experiment \cite{ZRus21}.
As a reminder, $l_{2\pi}$ is the distance where the phase of the signal generated in the following spatial slice is again constructive, i.e. the signal grows again. This corresponds to the thickness difference of two \mbox{minima}. In contrast to $l_c=l_{\pi}$ it is easier to directly extract $l_{2\pi}$ from experimental data, as the fast oscillations, which increase the uncertainty when determining the position of the extrema, are less pronounced for the minima. In the following, $l_{2\pi}$ will be computed, so that the results can be readily compared to experimental works\cite{ZRus21}. In general, $2l_{c}\neq l_{2\pi}$ and thus, only if the phase is a linear function of space, then $2l_{\pi}=l_{2\pi}$, which is not given for non-negligible focus conditions in the vicinity of the focus. 
In Fig.~\ref{fig:SHG-thickness-lambda-foc}(a) one now observes low frequency, but also high frequency oscillations, where the latter belong to the multi-reflections of the fundamental and SHG signal. The low frequency oscillations are due to the phase-mismatch and we see that the coherence length changes with wavelength as it is supposed to do with $n=n(\lambda)$. Considering the experimental values\cite{ZRus21} $l_{2\pi,exp}(850~nm)\approx 2.4$~µm and $l_{2\pi,exp}(950~nm)\approx 3.55$~µm, we can compare with the simulated data, as summarized in Tab.~\ref{tab:lc-comp-tab}.
 \begin{table}[]
     \centering
     \begin{tabular}{c|c|c}
          & $l_{2\pi}(850)$&   $l_{2\pi}(950)$\\
          \hline
          Experiment\cite{ZRus21}&(2.4$\pm0.1$)~µm &(3.55$\pm0.1$)~µm\\
          \hline
          Plane Wave&2.73~µm &4.03 ~µm\\
          \hline
          Approx. Eq.~\eqref{eq:approxnewcohl}&2.37~µm&3.36~µm\\
          \hline
          Full Model &2.42~µm &3.50~µm
     \end{tabular}
     \caption{Comparison between the experimental data of $l_{2\pi}$ for two selected wavelengths in the LN thin film and corresponding computed data, using the simulation data, as well as the approximation of Eq.~\eqref{eq:approxnewcohl} and the plane wave approximation.}
     \label{tab:lc-comp-tab}
 \end{table}
 The plane wave case delivers $l_{2\pi}(850~nm)\approx 2.73$~µm and $l_{2\pi}(950~nm)\approx 4.03$~µm which is in the same order of magnitude but over $10 \%$ to large.
 Using the approximation of Eq.~\eqref{eq:approxnewcohl} we obtain $l_{2\pi}(850~nm)\approx 2.37$~µm and $l_{2\pi}(950~nm)\approx 3.36$~µm, which is closer to the experimental value and only about $5\%$ off. The approximation works better for the lower wavelength, as the condition $L<b$ is better fulfilled. Finally, one can compute the exact solution, using the transcendental Eq.~\eqref{eq:exact-phi} (or extracting it from graph of the simulated data), yielding 
 $l_{2\pi}(850~nm)\approx 2.42$~µm and
 $l_{2\pi}(950~nm)\approx 3.50$~µm, respectively. These values are even closer and lie in the scope of the measurement uncertainty, which was determined to be $\pm 100$~nm in the corresponding paper\cite{ZRus21}. Beware that the experimental publication\cite{ZRus21} calls the observable $2l_c$, but it is in fact $l_{2\pi}$, the thickness difference between the first and second minimum. The actual calculation of $2 \cdot l_c=2\cdot l_{\pi}$ as twice the distance between the first minimum and maximum results in deviations, i.e. $2 \cdot l_c(850~nm)\approx 2.39$~µm and
 $2 \cdot l_c(950~nm)\approx 3.40$~µm. Here, the deviation increases with the coherence length, as the rate of phase-change, due to the Gouy-phase decreases with the distance, as the $arctan(x)$ converges to a constant. 
 The counter-propagating light on the other hand does again not play a significant role in this context, because the coherence length is orders of magnitude smaller and the co-propagating light is efficiently reflected from the substrate. Thus, the counter-propagating signal is significantly weaker.
 \par A detailed comparison of the coherent interaction lengths for different wavelengths and an $NA=0.45$ as a benchmark case is presented in Fig.~\ref{fig:lc-comp-ges}(a), where it is compared to experimental data and data sets of more rigorous numerical simulations\cite{ZRus21}. They are discussed in Sec.~\ref{cap:modelvsexp}.
 \subsubsection{Dependence for varying focus position}
 The former results used $z_{f}=0$, i.e. the focal plane is situated at the interface air/LN. In fact, for the case at hand, the overall trend of the signal does not change significantly, when changing the focal depth in a certain range. The resulting cross-section for SHG in a thin LN slab, which was simulated with the same parameters as before, is shown in Fig.~\ref{fig:SHG-thickness-lambda-foc}(b).
One can observe, that the oscillations are continued when focusing into the material, although, in detail, deviations exist. But these are very small in this respect due to $b\sim L$. Actually, the oscillation period is minimum, when the focus is placed at the center of the film ($z_{f}\approx t/2$, corresponding to a lower integral boundary of $z_{in}\approx -t/2$), while slightly growing when focusing closer to either of the interfaces. Indeed, this was also predicted in our previous work by full numerical calculations \cite{ZRus21}. Due to the large computational demand of those calculations, though, this could only be calculated for a few selected cases, while here a full picture can be discussed readily with the semi-analytical model.

\subsection{THG and the interference of different layers} \label{sec:thg-tf-ln}
This section focuses on the generation of third-harmonic light. As practically all materials, including substrates or the air above the sample, show a third order non-linearity, one acquires distinct results, that depend significantly more compared to SHG on the setup environment. As is illustrated in Fig.~\ref{fig:thg-var-os-ges}(c), for a typical microscopy setup, air (or oil/water) fills the space between lens and sample. As nonlinear optical microscopy operates with high pulse power, the air may as well interact with the light.
Thus, this additional layer has to be taken into account for the total picture of third-harmonic processes, too. Therefore another integral is added to the computation, featuring the corresponding boundaries, refractive indices, and a different $\chi^{(3)}$. To clearly see the interference effects, we chose the $\chi^{(3)}$ for the air to be of the same order of magnitude, i.e. as large as the one of the LN slab. \\
\subsubsection{The role of the $\chi^{(3)}$-relations}
\begin{figure}
    \centering
    \includegraphics[width=0.99\linewidth]{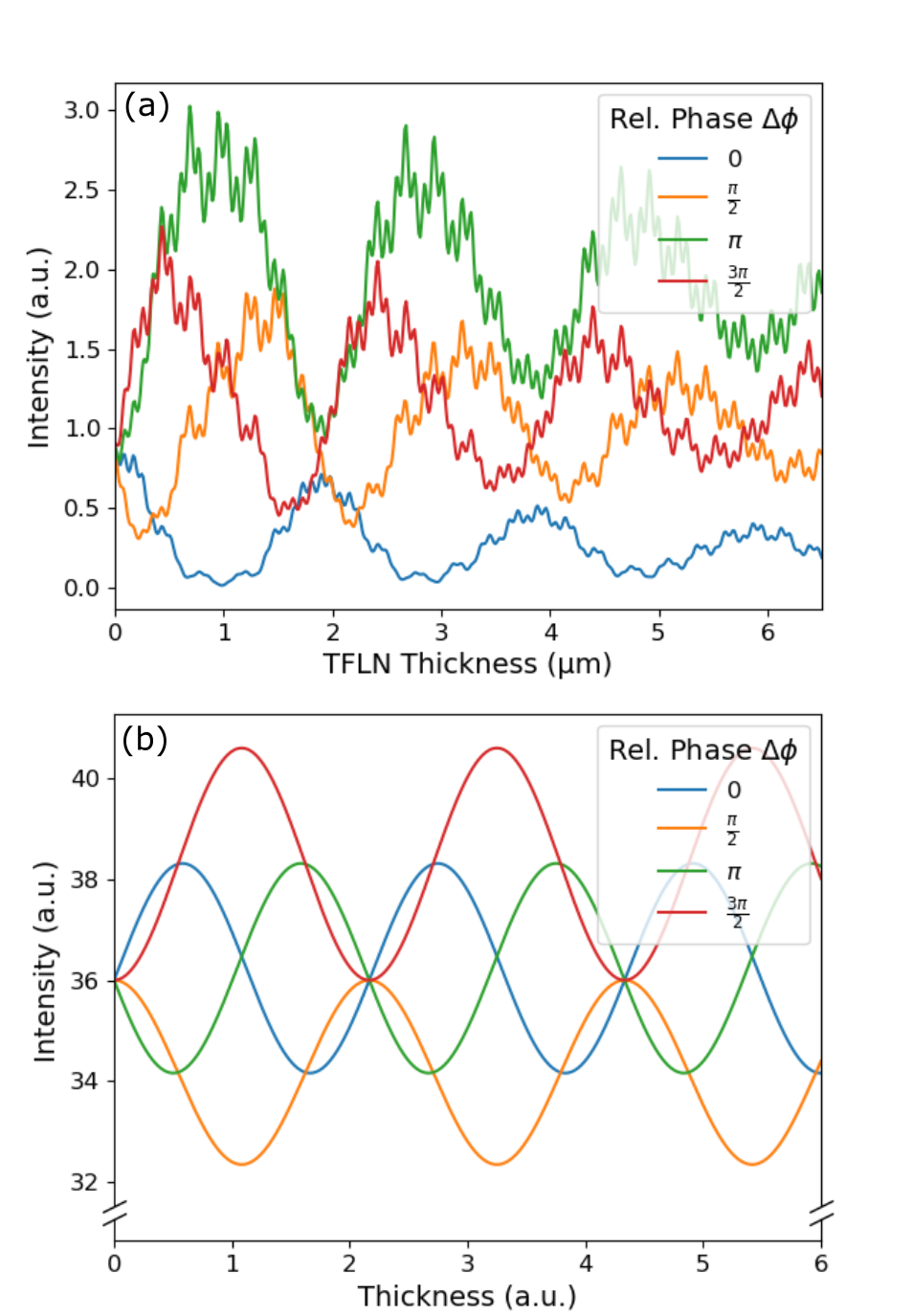}
    \caption{THG signal strength \textbf{(a)} in dependence of different phases between the $\chi^{(3)}$ of air and LN calculated with the Gaussian ansatz. The scan corresponds to a $z_{in}=0$ line-scan in Fig.~\ref{fig:thg-var-os-ges}(c). Here the amplitude of the susceptibility of air is equal to that of LN. The importance of the phase for the signal evolution is well observable; \textbf{(b)} shows an analogue calculation for the layer system in the plane wave model of Eq.~\eqref{eq:tmo-phase}. To obtain the (experimentally observed)\cite{ZRus22} results of the Gaussian beam simulations with no phase-shift, the air-signal needs to have a phase of $\pi/2$, i.e. the blue curve of \textbf{(a)} corresponds to the orange curve of \textbf{(b)}}
    \label{fig:thg-chivar-ges}
\end{figure}

For THG, the signal of the air above the surface can play a significant role. In air, signal is generated, where on one hand the phase-mismatch is very small due to the low dispersion while on the other hand, part of this generated light enters the LN slab and undergoes the same multireflections as the signal generated in the slab; together they generate a more complex interference pattern governed by the phase relations of the different $\chi^{(3)}$ values. As the nonlinear optical susceptibility can be a complex quantity in general, the result of interference patterns depends on the phase relation $\Delta \Phi = -i\log\left(\chi^{(3)}_{air}/ \chi^{(3)}_{LN}\right)$ of the contributing $\chi^{(3)}$-elements, cf. Fig.~\ref{fig:thg-chivar-ges}(a).
The data underlines, that the relative phase plays a significant role on the signal development. Although all signal curves show the characteristic low frequency oscillations, which stem from the phase-mismatch, especially the trend for small thicknesses is different. When the relation is a phase factor of $e^{0}=1$, the signal initially drops, as the newly generated third-harmonic signal of LN is seemingly out of phase; for $e^{\pi}=-1$ we get a growing signal. The signal drops or grows at first then, because the region, where LN-THG is generated, is growing, whereas its total phase is continously changing; therefore the interference pattern is shifted and also develops at small thicknesses at first. When the thickness reaches the size of the confocal parameter, the development becomes more stable, as less signal is generated in the air layer. Note that the figure shows the intensity and not the amplitudes; in the amplitudes, the oscillations of the zero and $\pi$ phase cases are (almost) symmetric. Interestingly, via the additional Gouy-phase, the susceptibilites with the same phase are (almost) destructively interfering at first.\\
This results demonstrate, that the amplitude and phase relation of the  nonlinear optical susceptibilities play an important role for the overall signal; This is as well true for the Gouy-phase. It does not only influence the coherence length, as already shown in different works\cite{Boyd92,ZRus21}, but also the phase of different interfering signals originating from different layers.
Here, our own work\cite{ZRus22} shows in fact that the case of no additional significant phase-shift of the non-linearities is realized in the thin film LN on silicon material system, i.e. the blue signal evolution of Fig.~\ref{fig:thg-chivar-ges}(a).
\subsubsection{Note on the phase evolution}In order to obtain a more intuitive and general picture of the influence of the air signal, one can compare the signal patterns for varying $\chi^{(3)}$ relations, as shown in Fig.~\ref{fig:thg-chivar-ges}, to a corresponding plane wave model and try to figure out different phase contributions. In the case of two interfering third-harmonic signals, the air signal has almost no phase mismatch, but presumably a collected Gouy-phase $\phi_1$, so that one obtains:
\begin{align}
    I&=\left|A\cdot e^{i\phi_1}+B\cdot \int_{0}^{L}e^{i\Delta k z} dz\right|^2,\; \text{with}\; A,B \in \mathcal{R_+}\\
    &=\left|A e^{i\phi_1}+ B \cdot e^{i\Delta k L/2}\frac{2}{\Delta k}\cdot \sin(\Delta k L/2)\right|^2\\
    &=|A|^2+|B \frac{2}{\Delta k} \sin(\Delta k L/2)|^2+ \nonumber \\
    &A B \frac{4}{\Delta k}\cos(\Delta k L/2 -\phi_1)\sin(\Delta k L/2)\;. \label{eq:tmo-phase}
\end{align}
Where in contrast to the plane wave case in one medium, one finds an additional background $|A|^2$ and an interference term $\propto \cos(\Delta k L/2 -\phi_1)$. Obviously, for vanishing $L$ only the background remains. However $\phi$ steers the further evolution; for non-focused systems, i.e. plane waves, one would expect no additional phase of the air signal and therefore a growing signal, similar to SHG with an additional background as shown in Fig.~\ref{fig:thg-chivar-ges}(b) in "blue", although the interference term leads to a phase shift.
For the focused case, an additional phase is added. If the focal plane lies for example on the interface,
the phase change 
can be extracted by computing the amplitude (and its phase) for the phase-matched beam in the air layer:
\begin{align}
    a_{air}\propto \int_{-\infty}^{0}\frac{1}{(1+i2z/b)^{2}}dz\\
    =\left[-\frac{b}{2i}\frac{1}{1+i2z/b}\right]_{-\infty}^{0}\\
    =-\frac{b}{2i}=\frac{b}{2}e^{i\frac{\pi}{2}}\\
    \Rightarrow \phi_1=\frac{\pi}{2}\;.
\end{align}
Obviously, the plane wave model with $\phi=\pi/2$ ("orange"), as in Fig.~\ref{fig:thg-chivar-ges}(a), fits perfectly to the case of no phase rotation of the susceptibility of the numerical analysis ("blue"), as in Fig.~\ref{fig:thg-chivar-ges}(b).\\
One can derive a corresponding relation for arbitrary harmonics and focal positions $z_0$, which is given in Eq.~\eqref{eq:airsig-phase-shift}:
\begin{align}
    a_{air,q}(z_0)&\propto \int_{-\infty}^{z_0}\frac{1}{(1+i2z/b)^{q-1}}dz\\
    &=\left[-\frac{b}{2i(q-2)}\frac{1}{(1+i2z/b)^{q-2}}\right]_{-\infty}^{z_0}\\
    &=\frac{b}{2(q-2)} \frac{e^{i\frac{\pi}{2}-i(q-2)arctan(2z_0/b)}}{\sqrt{1+(2z_0/b)^2}^{q-2}}\\
    &\Rightarrow \phi_1=\frac{\pi}{2}-(q-2)arctan(2z_0/b)\; . \label{eq:airsig-phase-shift}
\end{align}
One expects a phase that changes with focal position.
Noteworthy is also the fact, that the total phase is the same for all processes for $z_0=0$, although the higher harmonic beams pick up more  Gouy-phase. This is due to the higher suppression, i.e. lower intensity of the wavelets with larger phase, i.e. wavelets created far from the focal region have low intensity.\\
The equation does not necessarily hold for SHG. However, we have calculated the corresponding integral for the real part in Eq.~\eqref{eq:SHG-bulk-fin-signals-pm}; doing the same for the imaginary part with the integrand $\frac{4z/b}{1+(2z/b)^2}$ yields the stem function $\ln(1+(2z/b)^2)$, which diverges. The ratio of the imaginary to real part is therefore infinite and $arctan(x\rightarrow \infty)=\pi/2$, like for all the other harmonics.
\subsubsection{Third-harmonic signal for line-scans and cross-sections}
\begin{figure}
    \centering
    \includegraphics[width=0.99\linewidth]{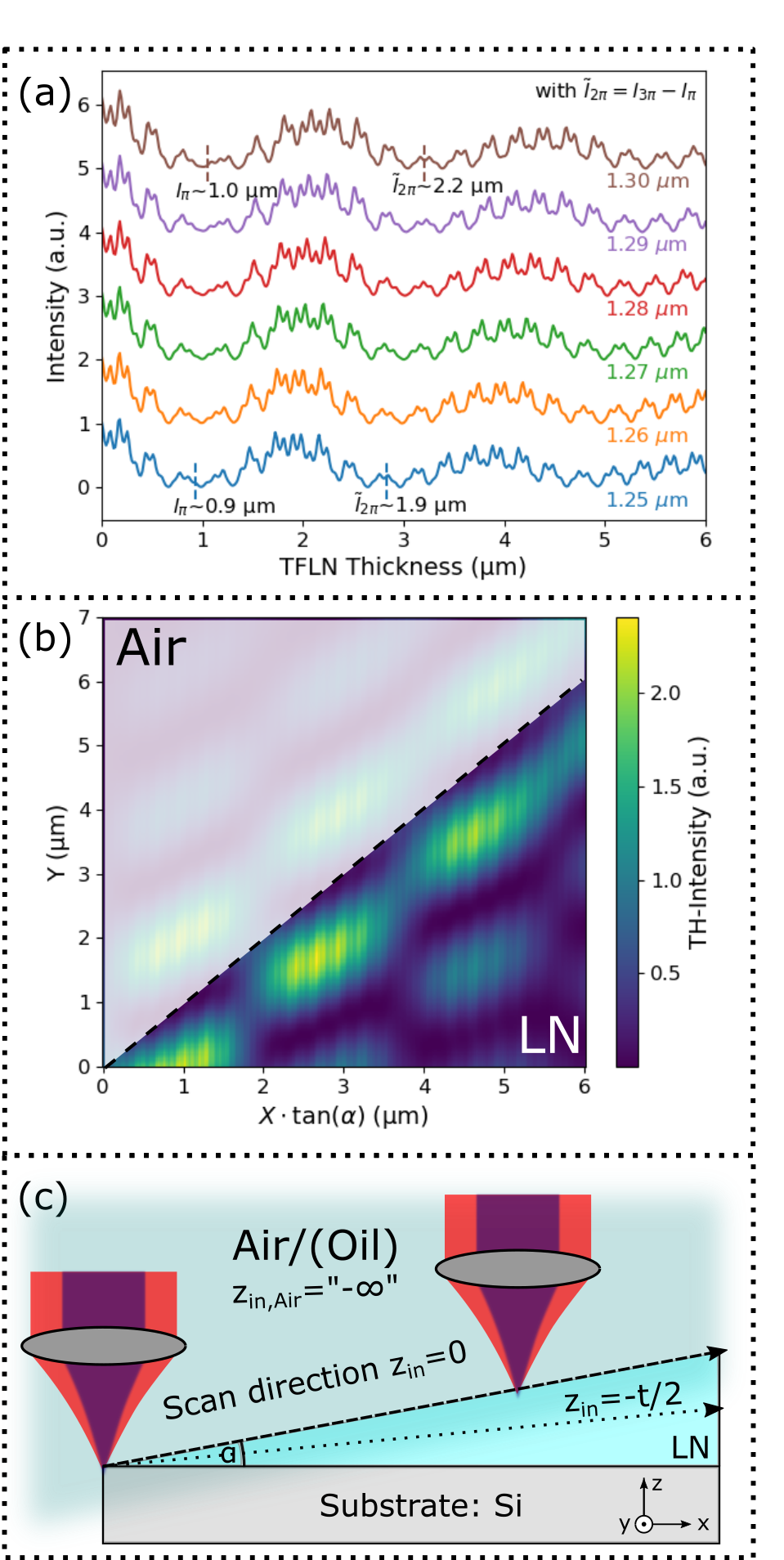}
    \caption{\textbf{(a)} Simulation of THG-signal for increasing material thickness for different fundamental wavelengths, corresponding to a line-scan on the surface ($z_{in}=0$) as depicted in (c). One observes an evolution of the coherence length which agrees well with  experimental data\cite{ZRus22}. The corresponding cross-section scan of the wedge \textbf{(b)} illustrates the dependence of the interference pattern on the focal depth. In \textbf{(c)} the measurement setup is sketched, i.e. for (a) a line-scan on the surface was carried out.}
    \label{fig:thg-var-os-ges}
\end{figure}

As expected, $l_c$ changes for the different wavelengths, as given by the dispersion of the material, cf. Fig.~\ref{fig:thg-var-os-ges}(a). One can also see quite well the interference with a $\phi_1=\pi/2$ signal due to air, as observed in the experiment\cite{ZRus22}.

Beware, that due to the different peak and dip distribution for THG, it is necessary to specify the distance, which is chosen to measure the coherent interaction length, since the rate of phase change and therefore the interference pattern is dependent on the thickness and focal position, as was shown in the last section. In fact $\tilde{l}_{2\pi}$, which is shown in Fig.~\ref{fig:thg-var-os-ges}(a), marks a phase difference of the wavelets of $2\pi$ relative to the first dip, which corresponds to a total phase of $\pi$, i.e. the marked minimum-to-minimum distance corresponds to $\tilde{l}_{2\pi}=l_{3\pi}-l_{\pi}$. It is in general larger than the preceeding maximum-to-maximum distance, i.e. $\tilde{l}_{2\pi}>l_{2 \pi}$; however, $\tilde{l}_{2\pi}$ can be directly compared with the equivalent minimum-to-minimum values of the experimental work\cite{ZRus22} of Amber et al.\\
A corresponding cross-section into the wedge, as depicted in Fig.~\ref{fig:thg-var-os-ges}(b) shows a significantly different signal pattern in contrast to SHG. \\
Since the total phase of the air-signal and the LN-signal change with the focal position, there is also a signal pattern in the depth-direction, which can be observed in Fig.~\ref{fig:thg-var-os-ges}(b). This can be seen in Eq.~\eqref{eq:airsig-phase-shift}, where the focal position as an argument of the $arctan(x)$ determines the phase of the air signal; equally, the translation of the focal position $z \rightarrow z-z_{foc}$ also shifts the integral boundaries of the LN signal. Approximating again with plane waves one obtains Eq.~\eqref{eq:zshift-vis}:
\begin{align}
    I&=\left|A\cdot e^{i\phi_1}+B\cdot \int_{0}^{L}e^{i\Delta k (z-z_{foc})} dz\right|^2,\; \text{with}\; A,B \in \mathcal{R_+}\\
    &=\left|A e^{i\phi_1}+B\cdot e^{i\Delta k L/2-i\Delta k z_{foc}}\frac{2}{\Delta k}*\sin(\Delta k L/2)\right|^2\\
    &=|A|^2+|B \frac{2}{\Delta k} \sin(\Delta k L/2)|^2+ \nonumber \\
    & A B \frac{4}{\Delta k}\cos(\Delta k L/2 - \Delta k z_{foc} -\phi_1)\sin(\Delta k L/2)\;. \label{eq:zshift-vis}
\end{align}
Thus, we obtain again a similar oscillation period in depth as for varying thickness, to which in principle both phase evolutions, i.e. of the signal generated in air and of the signal generated inside the material, contribute.\\
These are governed by the phase mismatch and the Gouy-phase contributions inside and outside the crystal. The z-dependent part of the argument of the cosine, shown in the approximated expression in Eq.~\eqref{eq:zshift-contri}, i.e.
\begin{align}
\phi_{depth}(z_{foc})&=\Delta k z_{foc}+\phi_1(b,z_{foc})\;,  \label{eq:zshift-contri}
\end{align}
allows for approximating the period as \mbox{$l_{depth, 2 \pi}\approx \frac{2 \pi}{\Delta k}+\mathcal{O}\left(\frac{1}{b}\right)$}. The similarity of both periods is visible in Fig.~\ref{fig:thg-var-os-ges}(b).

\subsection{Models vs. Experiment} \label{cap:modelvsexp}
\begin{figure}
    \centering
    \includegraphics[width=0.99\linewidth]{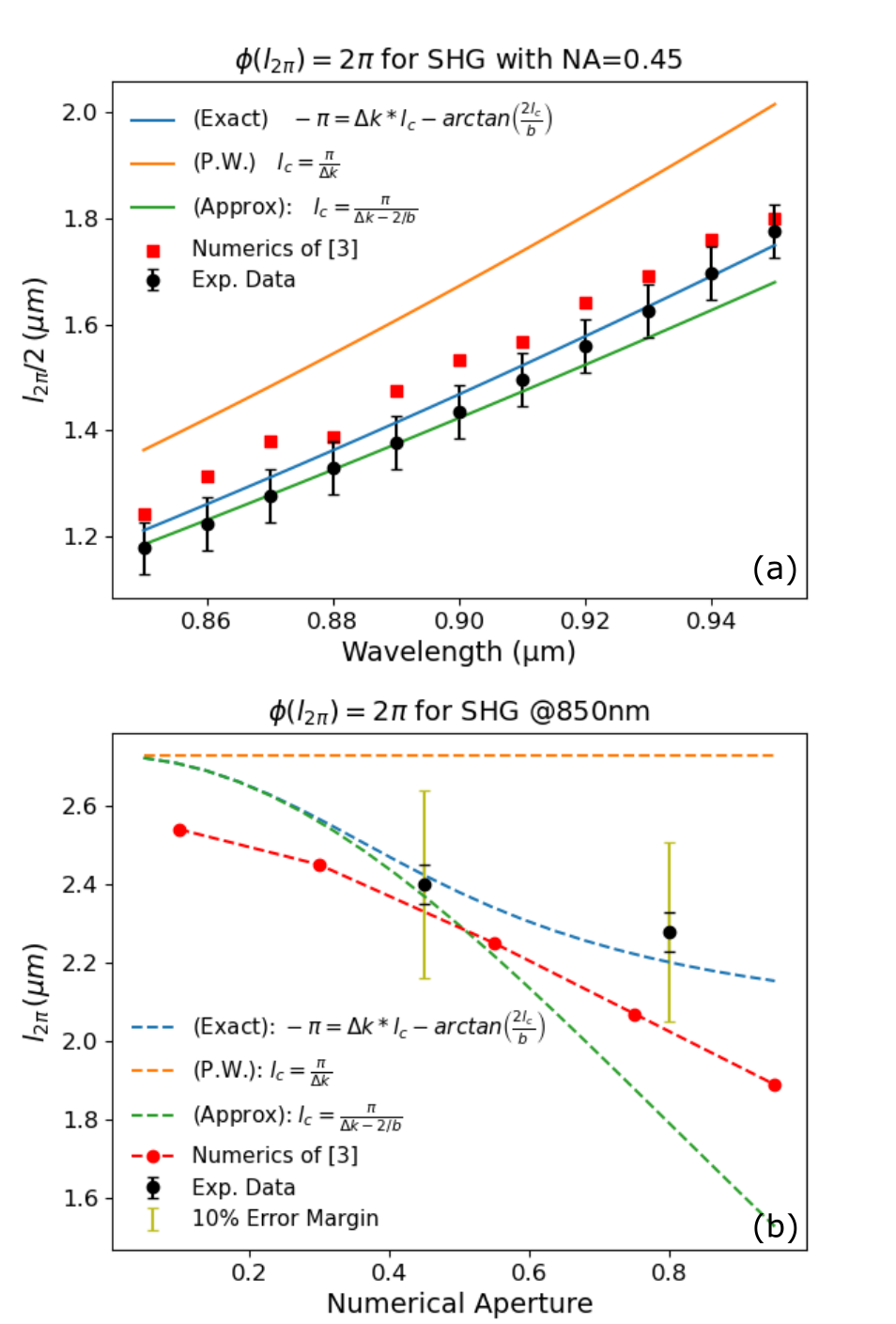}
    \caption{\textbf{(a)} Comparison of the  evolution of $l_{2\pi}$ for different calculation schemes with experimental data\cite{ZRus21}. Then in \textbf{(b)} the  evolution of $l_{2 \pi}$ for different NAs and calculation schemes are shown in relation to experimental data. For both, the full vectorial model \cite{ZRus21} is used besides the plane wave (P.W.) and the (semi-)analytical calculations of this paper, which can be compared to experimental values\cite{ZRus21}. Note that the legend shows how the corresponding model calculates  $l_c=l_{\pi}$.}
    \label{fig:lc-comp-ges}
\end{figure}

There is a quite rich experimental database for employing the modeling scenarios shown before. First of all, one can find a lot of depth scans done in the context of mapping ferroelectric domain walls, which correspond to the case of a half-infinite crystal. For example in past works\cite{Spy2020FM,Spy2017II} of the authors, one can find depth scans for high numerical apertures of $NA=0.95$, that show  the signal peak centered at the surface as calculated in Sec.~\ref{sec:shg-in-bulk} and depicted in Fig.~\ref{fig:surface-shg-ges} .\\
Then, there are second-harmonic measurements\cite{ZRus21} for wedged thin film lithium niobate on silicon, which agree very well with our simulations, which will be used in the following to compare the two models. One can also compare the depth-evolution as well as the changes of $l_c$ with the focusing conditions, cf. Fig.~\ref{fig:lc-comp-ges}(a) and Fig.~\ref{fig:lc-comp-ges}(b).


\begin{figure}
    \centering
    \includegraphics[width=0.97\linewidth]{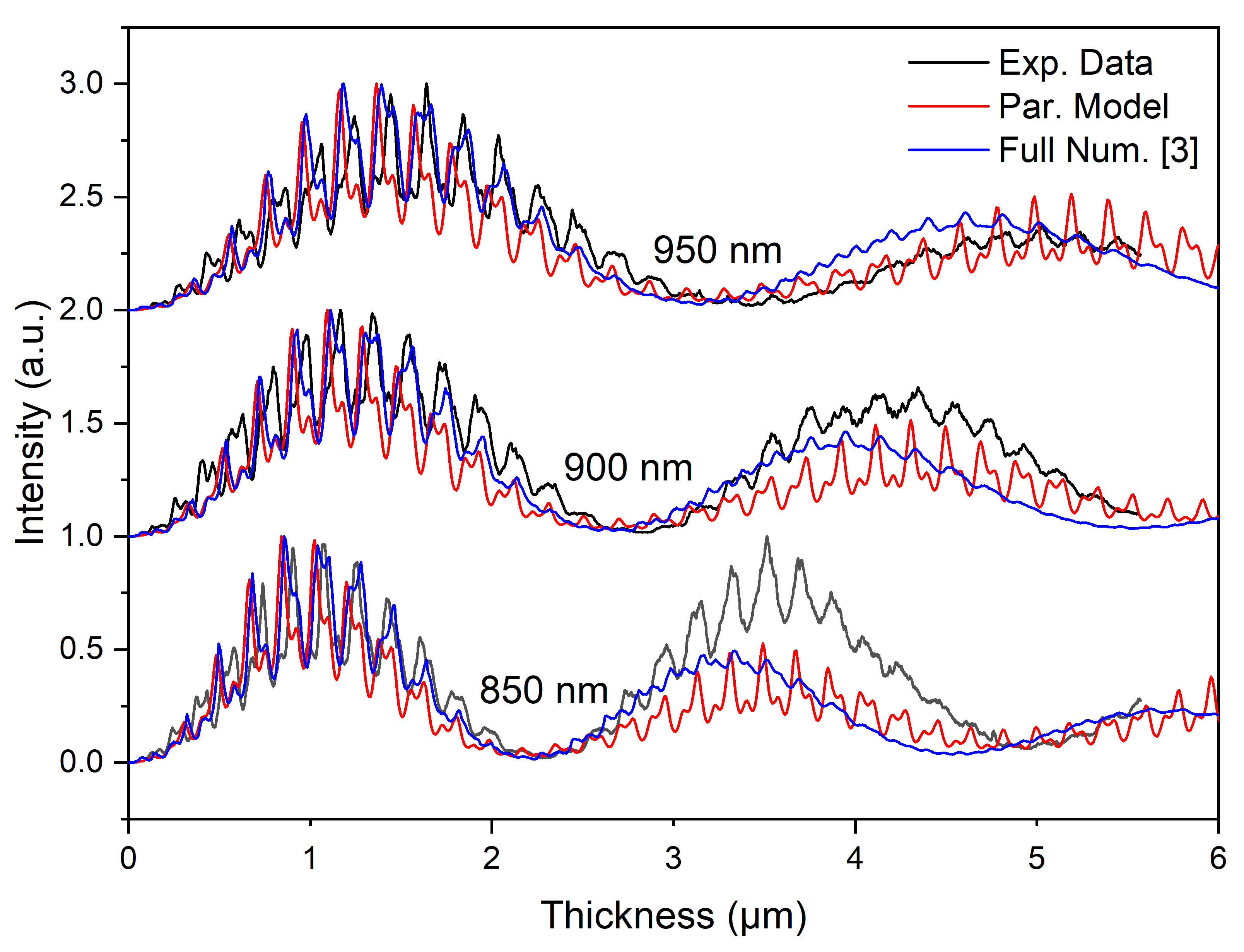}
    \caption{Second-harmonic signal data of line-scans, cf. Fig.~\ref{fig:thg-var-os-ges}(c), for different wavelengths using a NA of 0.8 together with the simulated thickness evolution using the full\cite{ZRus21} and the paraxial model.}
    \label{fig:exp-0NA8-SHG}
\end{figure}

In Fig.~\ref{fig:lc-comp-ges}(a), one observes an increasing match for the experimental $l_{2 \pi}$ and the simulated ones, when using the plane-wave, first-order-approximation and exact case, in that order. In comparison to the full simulation used in the source-paper, the effort for the calculation is much lower; that means several seconds for the full analysis on a typical personal computer for the semi-analytical model, since here only the transcendental equation \eqref{eq:exact-phi} has to be solved, as compared to at least several hours for the full numerical calculations for each wavelength.\\
Comparing the gathered data for different NAs, as depicted in  Fig.~\ref{fig:lc-comp-ges}(b), one can clearly observe the dependence of the  coherence-length, or more specifically $l_{2\pi}$ in this case, on the NA, which is not implemented in the plane-wave case.
As a note, the full 3D-numerics\cite{ZRus21} use a focal spot in the center of the layer, whereas the semi-analytical solutions are centered at the surface; in general, the extended models are much better suited as is the plane wave case, although for $NA>0.45$, the exact solution and the first order approximation diverge. Note that the paraxial approximation of the Gauss-beams becomes worse for high NAs, however as $n_{LN}\approx 2.2$, the effective NA is reduced, so that $\sin[\arcsin(0.8/2.2)]\approx 0.36\approx 0.8/2.2$ meaning $\sin(x)\approx x$ is valid and $\cos[\arcsin(0.8/2.2)]\approx 0.93$, so $\cos(x)\approx 1$ is roughly valid. That means that the paraxial approximation is still sufficiently satisfied, which can be illustrated when comparing the thickness-evolution of the signals in Fig.~\ref{fig:exp-0NA8-SHG}, which use a numerical aperture of $NA=0.8$ and yield a good agreement.
\section{Conclusion and Outlook}
In this work, we develop and discuss a semi-analytical modeling ansatz which can be used for simulation of nonlinear optical processes in stratified, layered, nonlinear media. It is based on an analytical ansatz\cite{Boyd92}, which solves the approximated nonlinear wave equation using Gaussian beams with spatially dependent amplitudes. This ansatz is re-casted to be suitable for finite bulk and (layered) thin film systems.
At the same time we try to conserve the instructive potential of the parameters of the analytical results.\\
\par
This concerns especially the need for a modified description of the coherent interaction length. It is shown, that it is dependent on the focus position and the actual total phase which is acquired, i.e. subsequent oscillations may have different period lengths. 
A corresponding mathematical formulation of the phase evolution is derived from the differential equation which follows from the Gaussian ansatz. Thus, it gets obvious that focusing is the reason for the re-interpretation, because the Gouy-phase of the fundamental beam introduces a non-linearity in the vicinity of the focus.
Therefore, a specific description of this parameters is necessary for comparison, i.e. relative position of the focus, oscillation order, and/or amount of total acquired phase. Only oscillations far away from the focal spot converge to the plane wave case of constant $l_c$, as the additional phase due to focusing, the Gouy-phase, tends to be a constant for large distances.\\
\par
In the second part, we apply the modeling ansatz to several example cases. The ansatz is applied for surface-near bulk scans as well as for thin layers on a  reflective substrate.\\
In the surface-near SHG bulk scans and for large phase-mismatches ($\Delta k \gg 1/b$), we observe second-harmonic signal, which is concentrated at the interface. Such surface-near SHG was already observed in several experiments\cite{Spy2017II,Spy2020FM}. Furthermore, for small phase-mismatches, the peak position of the second-harmonic signal can substantially ($\sim 1$~µm) deviate from the surface.\\
For thin film materials for SHG and THG we observe good agreement with rigorous 3D numerics\cite{Sandkuijl2013} and experiment\cite{ZRus21,ZRus22}. Especially for THG, the air layer also produces signal, which interferes with the signal of the nonlinear material, such that a specific interference pattern is visible. It is determined by the phase relations of both signals, which depends on the phase between their susceptibilities but also on the difference of acquired Gouy-phase in air and material.
\par
\par
Thus, the modeling ansatz can perform very well in comparison to rigorous numerics, and delivers an instructive notion of the contributing physical phenomena.
Furthermore, the calculations are very fast, i.e. in the order of a minute when using an ordinary personal computer device for a full simulation of several wavelengths, and even less for directly calculating specific parameters like the coherent interaction length.
\par
Concerning future research, our approach, or extensions of it, allow for the examination of further nonlinear optical processes besides harmonic generation, provided that they can be mapped to an adapted version of the ansatz. An interesting candidate for examination is the FWM process of Coherent Antistokes Raman Scattering (CARS), which constitutes an important tool for material characterization\cite{Reitzig:22}.
\par
In summary, although we point out the limits of the approach, which is bound to the form and parameter range of the analytical solution and thus cannot for example describe focus distortion\cite{Hell93,NasseWoehl:10}, it can be a practical complement to perform nonlinear optical analysis.
It opens up the possibility to easily identify the different effects which are at work, yielding an educational and easy-to-use toolkit for the presented experimental examinations, which facilitates to develop the setup and extract specific information from the experiment.

\section{Acknowledgements}
The authors gratefully acknowledge financial support by the Deutsche Forschungsgemeinschaft (DFG) through projects CRC1415 (ID: 417590517), EN 434/41-1 (TOP-ELEC), INST 269/656-1 FUGG and FOR5044 (ID: 426703838), as well as the Würzburg-Dresden Cluster of Excellence on “Complexity and Topology in Quantum Matter” - ct.qmat (EXC 2147; ID 39085490). Also, we would like to acknowledge the excellent support by the Light Microscopy Facility, a Core Facility of the CMCB Technology Platform at TU Dresden, where the SHG/THG analysis was performed.

\nocite{} 
\bibliography{aipsamp}

\end{document}